\documentclass[twocolumn,nofootinbib,floatfix,amsmath,amssymb,showpacs,superscriptaddress]{revtex4}

\usepackage{graphicx}
\usepackage{dcolumn}
\usepackage{bm}
\usepackage{color}

\begin{document}

\title{Low-lying $\Lambda$ Baryons with spin 1/2 in Two-flavor Lattice QCD}

\author{Toru T. Takahashi}
\affiliation{Yukawa Institute for Theoretical Physics, Kyoto University,
Sakyo, Kyoto 606-8502, Japan}
\author{Makoto Oka}
\affiliation{Department of Physics, H-27, Tokyo Institute of Technology, Meguro, Tokyo 152-8551 Japan}

\date{\today}

\begin{abstract}
Low-lying $\Lambda$ baryons with spin 1/2 are analyzed in full (unquenched) lattice QCD.
We construct $2 \times 2$ cross correlators 
from flavor SU(3) ``octet'' and ``singlet'' baryon operators,
and diagonalize them so as to extract information of 
two low-lying states for each parity.
The two-flavor CP-PACS gauge configurations are used, which are generated
in the renormalization-group improved gauge action
and the ${\mathcal O}(a)$-improved quark action.
Three different $\beta$'s,
$\beta = 1.80$, 1.95 and 2.10, are employed, 
whose corresponding lattice spacings are $a = 0.2150$, 0.1555 and 0.1076 fm. 
For each cutoff, we use four hopping parameters,
($\kappa_{\rm val}, \kappa_{\rm sea}$), which correspond to the pion masses
ranging about from 500 MeV to 1.1 GeV.
Results indicate that there are two negative-parity $\Lambda$ states nearly degenerate 
at around 1.6 GeV,
while no state as low as $\Lambda (1405)$ is observed.
By decomposing the flavor components of each state, 
we find that the lowest (1st-excited) negative-parity state
is dominated by flavor-singlet (flavor-octet) component.
We also discuss meson-baryon components of each state,
which has drawn considerable attention
in the context of multi-quark pictures of $\Lambda (1405)$.
\end{abstract}
\maketitle

\section{Introduction}

$\Lambda (1405)$ is one of the most interesting hadrons 
and attracting much interest from several view points. 
$\Lambda (1405)$ is the lightest negative-parity baryon, 
even though it has one valence strange quark, 
which is heavier than the up and down quarks.
Among the $J^P = 1/2^-$ baryons, $\Lambda (1405)$ is 
isolated from the others,
much lighter than
the non-strange counterpart $N(1535)$.
It has no spin-orbit partner in the vicinity, as the lowest spin $3/2^-$
state is $\Lambda (1520)$.
Furthermore, the structure of $\Lambda (1405)$ remains mysterious. 
On one hand, $\Lambda (1405)$ is interpreted as a flavor-SU(3)-singlet 
three-quark state in conventional quark models.
On the other hand, 
$\Lambda (1405)$ could be interpreted as an antikaon-nucleon 
$\bar KN$ molecular bound state (B.E. $\sim$ 30 MeV). 
The binding energy of $\bar KN$ implies a
strong attractive force between $\bar K$ and $N$
~\cite{Sakurai:1960ju,Dalitz:1967fp}, 
which may cause a new type of dense hadronic
matter, kaonic nuclei or kaonic nuclear matter
~\cite{Akaishi:2002bg,Yamazaki:2002uh,Akaishi:2005sn}.
We also expect that  such $\bar KN$ bound states with large
binding energies can be regarded as compact 5-quark states.
The 5-quark picture of $\Lambda(1405)$ has advantages
that all five quarks can be placed in the lowest-lying $L=0$ state
to form a negative-parity baryon, and also 
that it requires no spin-orbit partner of $\Lambda (1405)$.

The property of $\Lambda (1405)$ 
can therefore be an important clue to new paradigm in hadron physics. 
We here study properties of $\Lambda(1405)$ using the lattice QCD formulation.
Lattice QCD is nowadays employed as a very 
powerful tool for nonperturbative analysis directly based on QCD,
and expected to cast light on the nature of $\Lambda (1405)$ 
in a model-independent way. 
Though several lattice QCD studies
on $\Lambda (1405)$ have been performed so far
~\cite{Melnitchouk:2002eg,Nemoto:2003ft,Burch:2006cc,Ishii:2007ym},
most of them are based on quenched QCD and 
few lattice QCD studies succeeded in reproducing the mass of $\Lambda (1405)$.
Moreover, little has been discussed on the lattice
about the possible mixing
of flavor-SU(3)-octet and -singlet components induced by
the symmetry breaking.

Several possible reasons for the failure of reproducing $\Lambda (1405)$ in lattice QCD
were suggested through these studies, such as
missing meson-baryon components due to quenching,
exotic (non-3-quark type) structure of $\Lambda (1405)$,
or insufficiency of the lattice volume in the simulations.
Resolving such difficulties requires 
unquenched lattice QCD calculation on a larger lattice volume
with varieties of interpolating operators.

In this paper, 
we aim at clarifying the properties of $\Lambda (1405)$
with two-flavor full lattice QCD, adopting the ``octet'' and ``singlet''
baryon operators to construct correlation matrices,
which enables us to extract the low-lying spectrum
as well as the mixing between octet and singles
components in $\Lambda (1405)$.

\section{Lattice QCD setups}

\subsection{simulation conditions}

We adopt the renormalization-group improved gauge action
and the ${\mathcal O}(a)$-improved quark action.
Simulation parameters are listed in Table~\ref{simparam}.
We adopt three different $\beta$'s,
$\beta = 1.80$, 1.95 and 2.10,
and corresponding lattice spacings are $a = 0.2150$, 0.1555 and 0.1076
fm~\cite{AliKhan:2001tx},
which are determined so that the empirical $\rho$-meson mass is reproduced.
The hopping parameters for strange quark $\kappa_s$
are set to be 0.1431, 0.1393, and 0.1373 at $\beta$=1.80, 1.95, and 2.10
so that the Kaon mass is reproduced.
We note here that, if we adopt $\phi$-meson mass as an input,
$\kappa_s$ would be slightly different from those determined by the Kaon mass.
The dynamics of $\Lambda$ resonances would be closely connected
to Kaon dynamics, and hence we adopt Kaon mass as an input in this study.
We employ four different hopping parameters 
($\kappa_{\rm val}, \kappa_{\rm sea}$)
for each cutoff.
Corresponding pion masses range approximately from 500MeV to 1.1 GeV
at each $\beta$.

\begin{table}[h]
\begin{tabular}{p{0.05\textwidth}p{0.05\textwidth}p{0.1\textwidth}p{0.08\textwidth}p{0.08\textwidth}c}
\hline
$\beta$ & $c_{\rm SW}$ & $N_s^3\times N_t$ & $a$ [fm] & $N_sa$ [fm] & $\kappa_s$ \\ \hline\hline
1.80 & 1.60 & $12^3\times 24$ & 0.2150 & 2.580 & 0.1431\\ 
1.95 & 1.53 & $16^3\times 32$ & 0.1555 & 2.488 & 0.1393\\
2.10 & 1.47 & $24^3\times 48$ & 0.1076 & 2.582 & 0.1373\\ \hline
\end{tabular}
\caption{\label{simparam}
List of the present simulation parameters.
$\beta$ and $c_{\rm SW}$ are the parameters for the renormalization-group
improved gauge action and the ${\mathcal O}(a)$-improved quark action.
$N_s$ and $N_t$ denote the spatial and temporal lattice extents,
and the lattice spacings $a$ and the strange-quark hopping parameters
$\kappa_s$ are determined so that the empirical $\rho$-meson
mass and Kaon mass are reproduced~\cite{AliKhan:2001tx}.
}
\end{table}

\subsection{baryonic operators}

In order to extract the low-lying states
in $S=-1$ and isosinglet channel,
we construct $2\times 2$ cross correlators
from the following ``singlet'' and ``octet'' operators,
\begin{eqnarray}
{\eta}_{\bf 1}(x)
\equiv
\Lambda_1(x)
=
\frac{1}{\sqrt{3}}
\epsilon^{abc}
\left\{
u^a(x)[d^{T b}(x) C \gamma_5 s^c(x)] 
\right. \nonumber \\ \left.
+
d^a(x)[s^{T b}(x) C \gamma_5 u^c(x)]
+
s^a(x)[u^{T b}(x) C \gamma_5 d^c(x)] 
\right\}
\label{defeta1}
\end{eqnarray}
\begin{eqnarray}
{\eta}_{\bf 8}(x)
\equiv
\Lambda_8(x)
=
\frac{1}{\sqrt{6}}
\epsilon^{abc}
\left\{
u^a(x)[d^{T b}(x) C \gamma_5 s^c(x)]
\right. \nonumber \\ \left.
+
d^a(x)[s^{T b}(x) C \gamma_5 u^c(x)]
-2
s^a(x)[u^{T b}(x) C \gamma_5 d^c(x)]
\right\}
\label{defeta2}
\end{eqnarray}
It is easy to check that ${\eta}_{\bf 1}(x)$
(${\eta}_{\bf 8}(x)$) belongs to the singlet (octet) 
irreducible representation of the flavor SU(3).
The cross correlator of ${\eta}_{\bf 1}$ and  ${\eta}_{\bf 8}$ 
vanishes in the flavor-SU(3) symmetric limit ($m_u=m_d=m_s$).
We define cross correlators as
\begin{equation}
{\mathcal M}(x,y)_{IJ}
\equiv
{\rm Tr}
\frac{1\pm\gamma_4}{2}
\langle
{\eta}_I(x)
\bar{\eta}_J(y)
\rangle,
\end{equation}
where the trace is taken over spinor indices.
The parity-projection operator, 
$P(\pm) \equiv \frac{1\pm\gamma_4}{2}$, is inserted 
to separate the positive and negative parity states.
In the actual simulation, 
we adopt point-type operators for the sink, $\eta(x)$,
and extended operators, which are smeared
in a gauge-invariant manner, for the source, $\bar\eta(y=0)$.
Smearing parameters are chosen so that 
root-mean-square radius is approximately 0.5 fm.

\subsection{diagonalization of cross correlators}

We here consider a general situation 
where we have a set of $N$ independent operators,
$\eta^{\rm snk}_I$ for sinks and $\eta^{\rm src \dagger}_I$ for
sources in order to construct correlation matrices
${\mathcal M}_{IJ}(t)\equiv\langle \eta^{\rm snk}_I(t)
\eta^{\rm src \dagger}_J(0)\rangle$,
which can be
decomposed into the sum over the energy eigenstates $|i \rangle$ as
\begin{eqnarray}
{\mathcal M}_{IJ}(t)
&\equiv&
{\mathcal M}_{IJ}(t,0)
=
\langle \eta^{\rm snk}_I(t) \eta^{\rm src \dagger}_J(0)\rangle
\nonumber \\
&=&
\sum_{i,j}
(C^\dagger_{{\rm snk}})_{Ii}
\Lambda(t)_{ij}
(C_{{\rm src}})_{jJ} \nonumber \\
&=&
(C^\dagger_{\rm snk} \Lambda(t) C_{\rm src})_{IJ},
\end{eqnarray}
where the small letters ($ij$)
are the indices for the intermediate
mass eigenstates and
\begin{eqnarray}
(C^\dagger_{\rm snk})_{Ii}\equiv \langle {\rm vac} | \eta^{\rm snk}_I | i \rangle, 
\quad
(C_{\rm src})_{jI}\equiv \langle j | \eta^{\rm src \dagger}_J | {\rm vac} \rangle, 
\end{eqnarray}
are the coupling matrices.
$\Lambda(t)$ is a diagonal matrix given by the 
energy eigenvalue, $E_i$, of the state $i$.
\begin{equation}
\Lambda(t)_{ij}\equiv \delta_{ij} e^{- E_it}.
\label{dmatrix}
\end{equation}
From the product 
\begin{equation}
{\mathcal M}^{-1}(t+1){\mathcal M}(t)
=C_{\rm src}^{-1}\Lambda(-1)C_{\rm src},
\end{equation}
we extract the effective eigen-energies 
(or the effective masses after the zero-momentum projection)
\{$E_i$\} 
at the time slice $t$
as
the logarithm of eigenvalues \{$e^{E_i}$\} of the matrix 
${\mathcal M}^{-1}(t+1){\mathcal M}(t)$.

Besides overall constants,
$(C_{\rm src})^{-1}$ and $(C^\dagger_{\rm snk})^{-1}$
can be obtained as right and left eigenvectors
of 
${\mathcal M}^{-1}(t+1){\mathcal M}(t)$ and ${\mathcal M}(t){\mathcal M}(t+1)^{-1}$
respectively,
since
\begin{equation}
{\mathcal M}^{-1}(t+1){\mathcal M}(t)(C_{\rm src})^{-1}
=(C_{\rm src})^{-1}\Lambda(-1)
\end{equation}
and
\begin{equation}
(C^\dagger_{\rm snk})^{-1}{\mathcal M}(t){\mathcal M}(t+1)^{-1}
=\Lambda(-1)(C^\dagger_{\rm snk})^{-1}
\end{equation}
hold.
One can also directly construct optimal source and sink operators,
${\cal O}^{\rm src \dagger}_i$ and 
${\cal O}^{\rm snk}_i$,
which couple dominantly (solely in the ideal case) to $i$-th lowest state,
as
\begin{equation}
{\cal O}^{\rm src \dagger}_i=\sum_J
\eta_J^{\rm src \dagger} (C_{\rm src})^{-1}_{Ji}
\end{equation}
and
\begin{equation}
{\cal O}^{\rm snk}_i=\sum_J
(C^\dagger_{\rm snk})^{-1}_{iJ} \eta_J^{\rm snk}.
\end{equation}
When an optimal correlator 
which corresponds to a single-state propagation is needed,
one can extract it by sandwiching ${\mathcal M}(t)$ between
$(C^\dagger_{\rm snk})^{-1}$ and $(C_{\rm src})^{-1}$
since 
\begin{equation}
(C^\dagger_{\rm snk})^{-1} {\mathcal M}(t)(C_{\rm src})^{-1}
=\Lambda(t)
\label{diag01}
\end{equation}
is diagonal.
We note here that,
if the correlation matrix ${\mathcal M}(t)$ is hermitian,
one can determine
$(C_{\rm src})^{-1}$ and $(C_{\rm snk})^{-1}$ up to overall phase factors
so that Eq.(\ref{diag01}) is satisfied.

These prescriptions are valid only in the $t$-range where the number of 
the relevant physical states is no more than $N$,
since we prepare a set of just $N$ independent operators.
In practice, higher excited states may come in,
and such contamination 
manifests itself as a $t$-dependence
in several ``constants'' mentioned above.
In order to avoid the contamination of the excited states, 
we identify a $t$-window in which the ``constants'' are 
(almost) $t$-independent.
All the physical quantities are extracted within such a $t$-window.

\subsection{different choices in formulation}

We here discuss the difference in formulation.
In general, we solve generalized eigenvalue problems~\cite{Luscher:1990ck,Perantonis:1990dy},
\begin{eqnarray}
&&{\mathcal M}^{-1}(t'){\mathcal M}(t)
=(C_{\rm src}^{-1})\Lambda(t-t')(C_{\rm src}), \\
&&{\mathcal M}(t){\mathcal M}^{-1}(t')
=(C^\dagger_{\rm snk})\Lambda(t-t')(C^\dagger_{\rm snk})^{-1},
\end{eqnarray}
where we have ambiguity in the choice of $t'$.
As long as a system is dominated by the states as many as the rank of
correlation matrices, choice dependence vanishes.
In practice, we may have excited-state contaminations,
and choice dependence can emerge.
In our case, $|t-t'|$ is fixed to 1, and we take as large $t$ as possible:
We seek plateau regions and fit data in that window.
The authors in Ref.~\cite{Blossier:2009kd} investigated
possible subleading contaminations by higher excited states 
and the dependence on the choice of $t$ and $t'$.

\begin{figure}[hbt]
\includegraphics[scale=0.35]{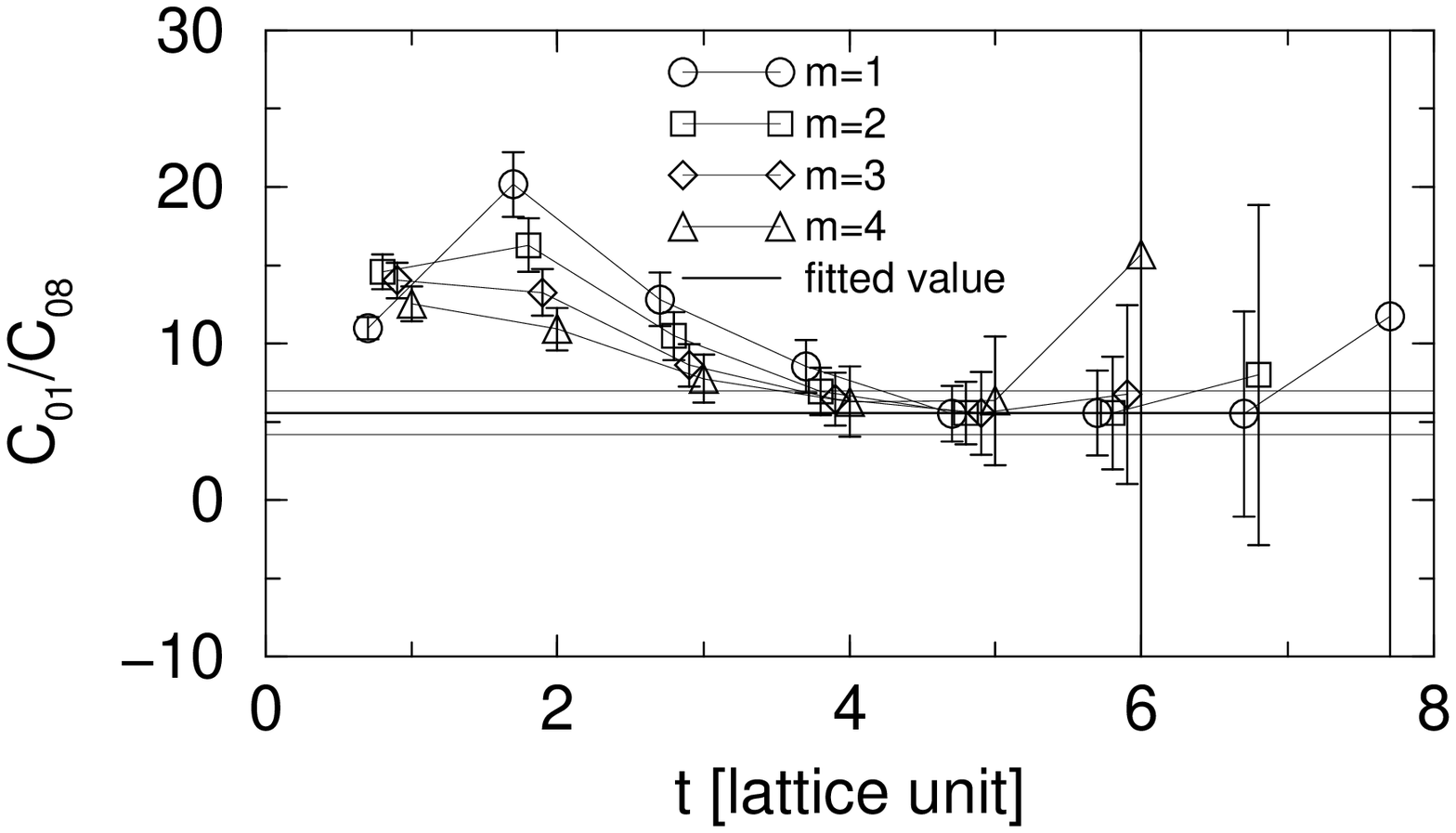}
\includegraphics[scale=0.35]{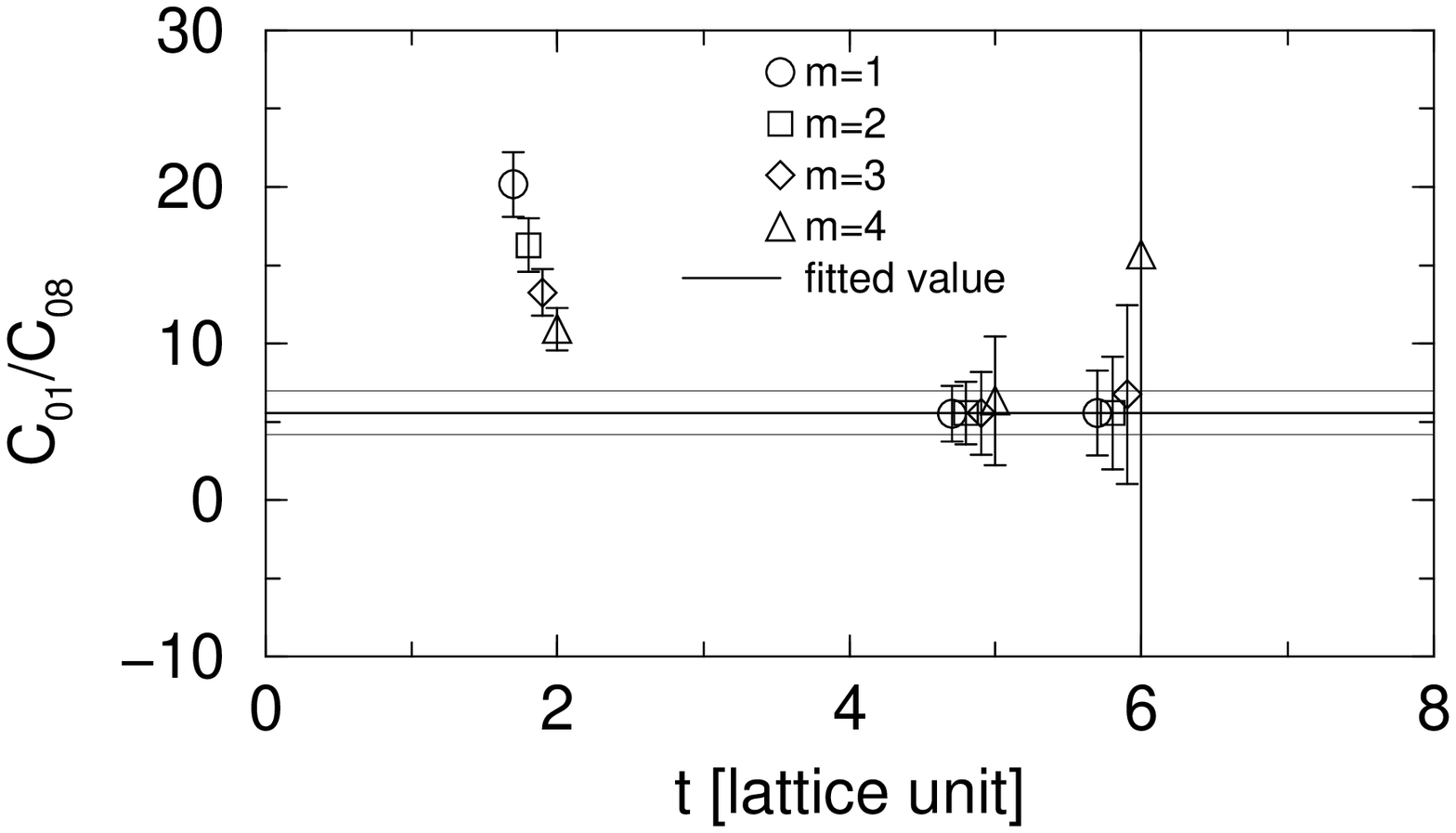}
\caption{\label{tcheckexample}
$(C_{\rm snk})_{0{\bf 1}}$/$(C_{\rm snk})_{0{\bf 8}}$
obtained by solving eigenvalue problem,
${\mathcal M}(t){\mathcal M}^{-1}(t+m)
=(C^\dagger_{\rm snk})\Lambda(-m)(C^\dagger_{\rm snk})^{-1}$
with $m\leq 4$ are plotted as a function of $t$.
They are obtained at
$\kappa_{\rm sea}, \kappa_{\rm val} = 0.1367$ at $\beta = 2.10$.
The horizontal lines denote the fitted value and the corresponding error
evaluated by fitting plateaus.}
\end{figure}

To estimate the systematic errors coming from such ambiguity,
we try four different choices,
$|t-t'|$=1, 2, 3, and 4, and see the dependence.
Namely, we solve the eigenvalue problems
\begin{equation}
{\mathcal M}(t){\mathcal M}^{-1}(t+m)
=(C^\dagger_{\rm snk})\Lambda(-m)(C^\dagger_{\rm snk})^{-1},
\end{equation}
with $m\leq 4$.
We plot in Fig.~\ref{tcheckexample}
$(C_{\rm snk})_{0{\bf 1}}$/$(C_{\rm snk})_{0{\bf 8}}$,
which can be a measure of flavor contents in an eigenstate 
and will be discussed later.
They are obtained at
$\kappa_{\rm sea}, \kappa_{\rm val} = 0.1367$ at $\beta = 2.10$.
Though they are all t-independent in an ideal case,
there remain considerable contaminations especially with small $m$.
On the other hand, it reaches plateau region much faster with larger $m$.
However larger $m$ results in larger statistical fluctuations
and diagonalization sometimes fails at relatively small $t$.
In our case, as long as we identify plateaus properly,
resulting values are all consistent with each other.

\section{Lattice QCD Results}

\subsection{hadronic masses}

\begin{figure}[hbt]
\includegraphics[scale=0.35]{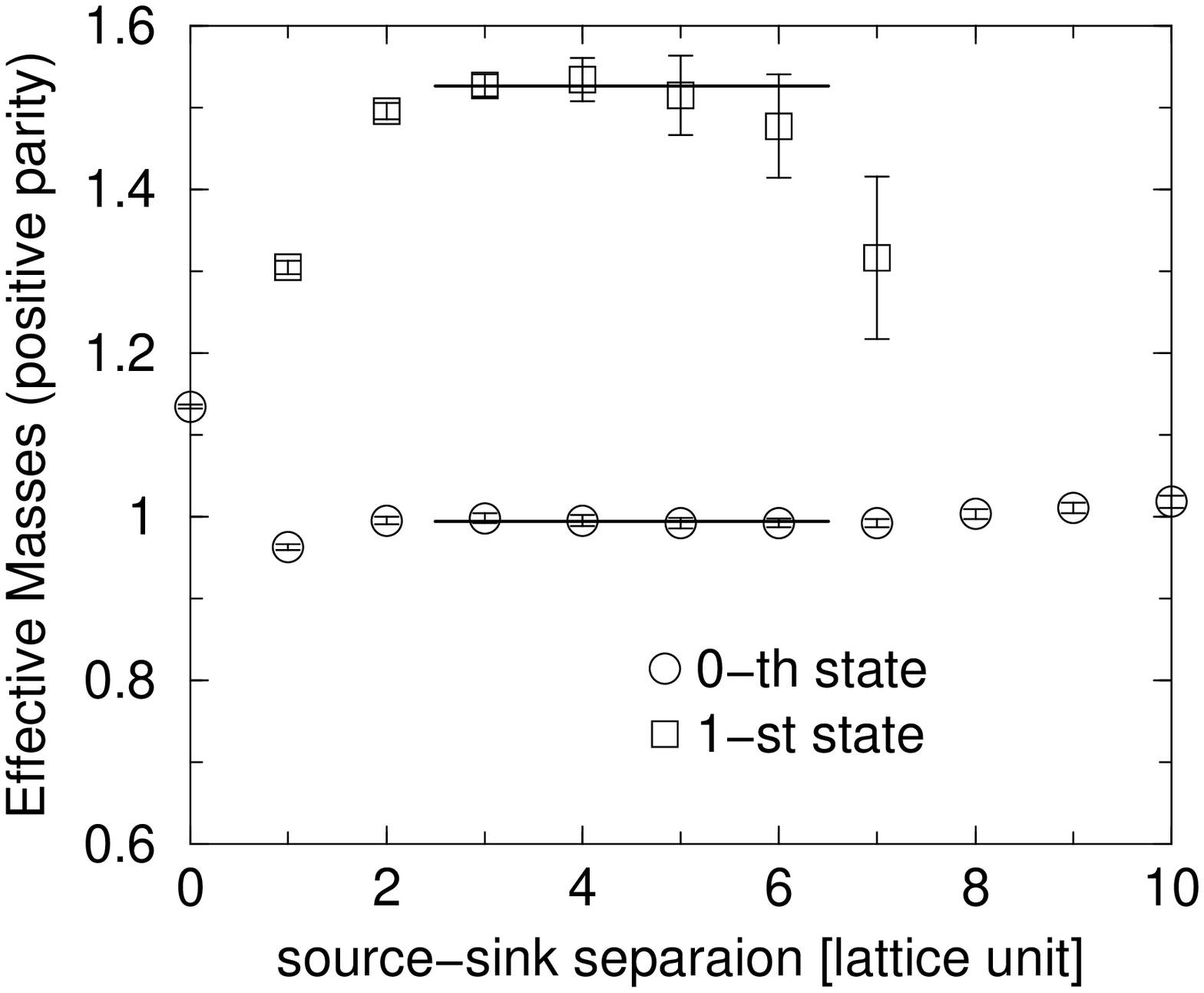}
\includegraphics[scale=0.35]{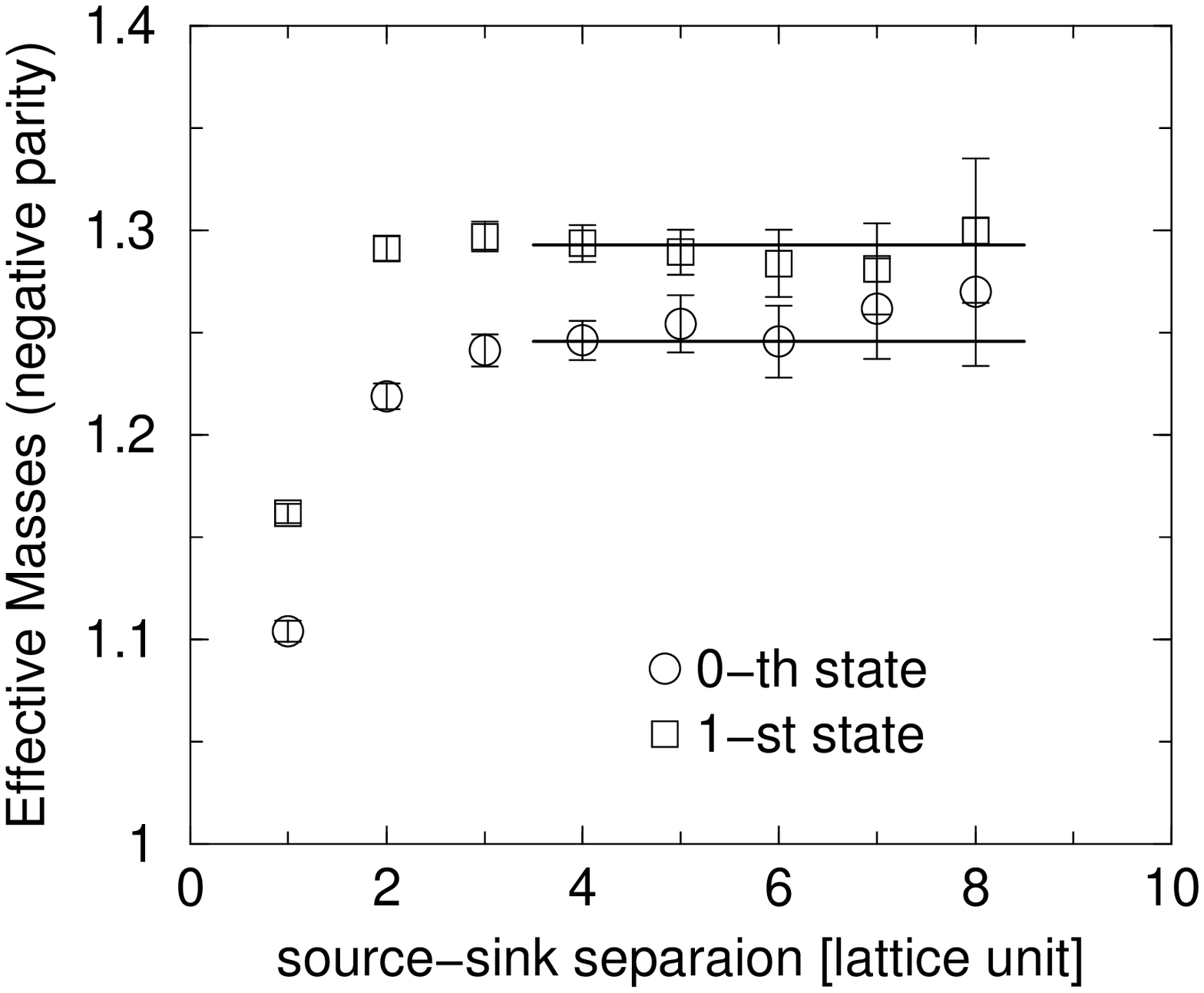}
\caption{\label{effmassexample}The effective mass plots
for ground and 1st-excited states in the positive- (upper panel) 
and negative-parity (lower panel)
channels, which are obtained with 
$\kappa_{\rm sea}, \kappa_{\rm val} = 0.1367$ at $\beta = 2.10$.
The horizontal lines are fits of the eigen-energies in the plateau region
for each state.}
\end{figure}

\begin{figure}[hbt]
\includegraphics[scale=0.35]{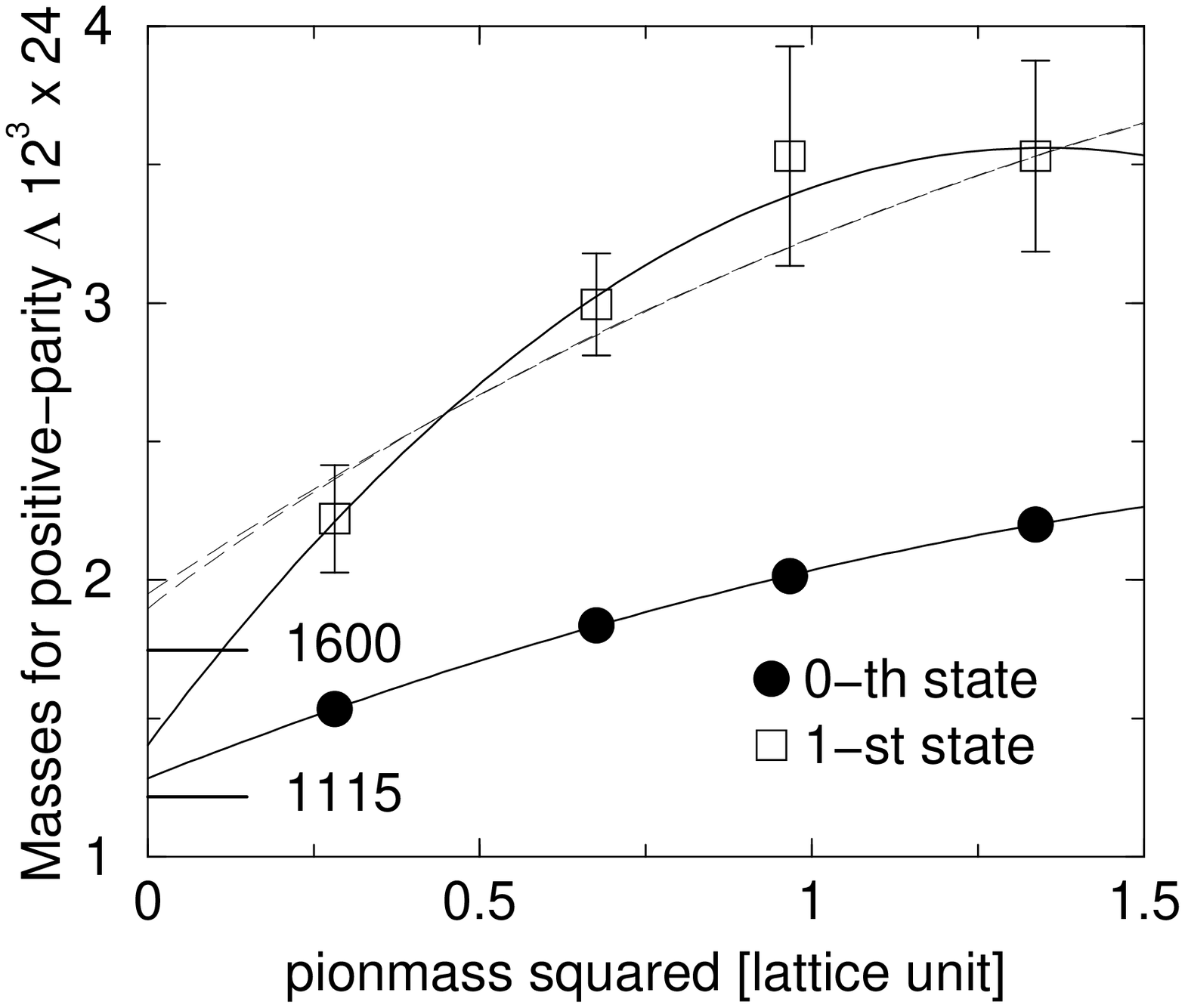}
\includegraphics[scale=0.35]{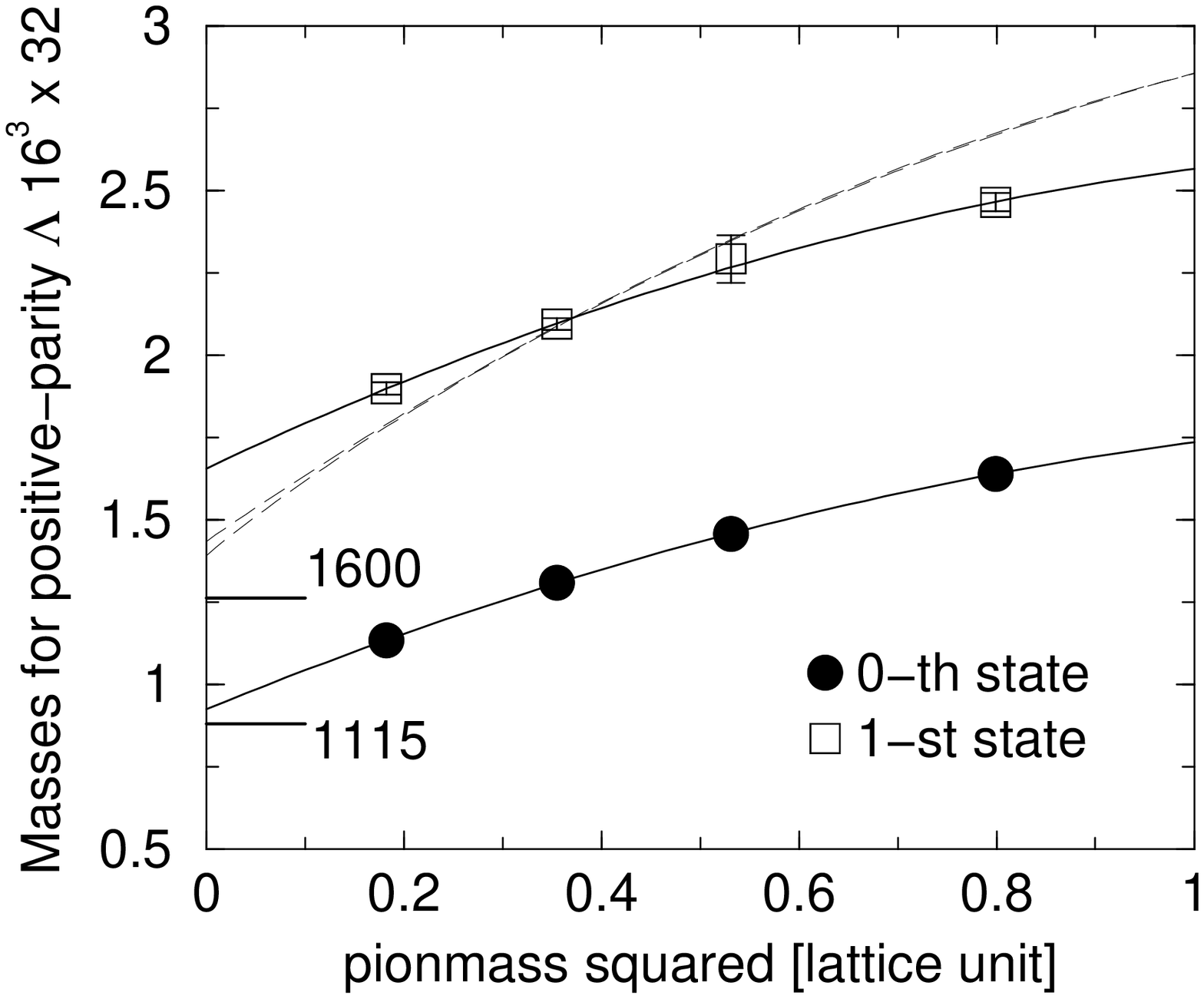}
\includegraphics[scale=0.35]{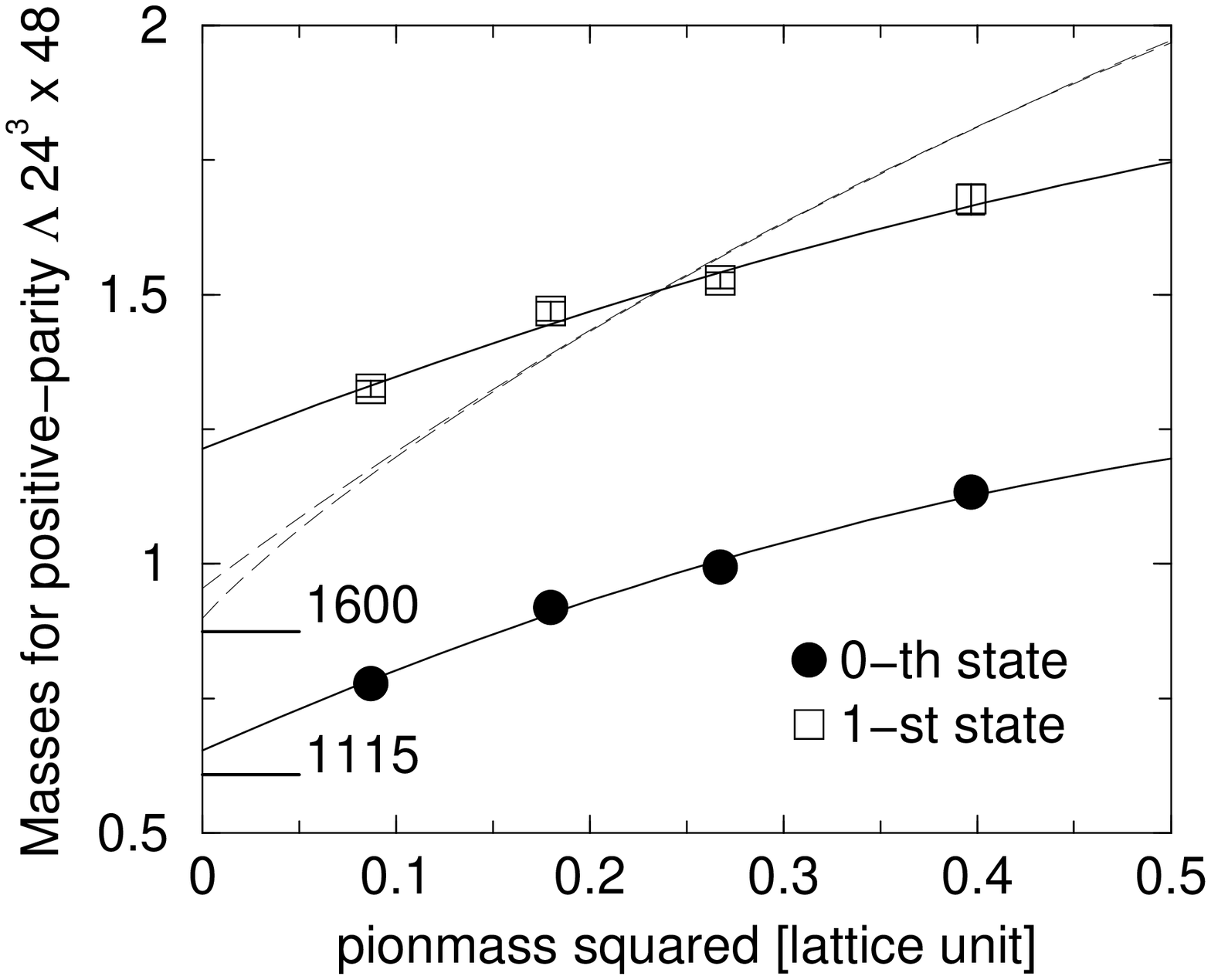}
\caption{\label{posmass}
Masses of the lowest two positive-parity $\Lambda$ states
as functions
of the squared pion mass. 
The filled circles (open squares) denote
the masses of lowest (1st-excited) state.
Two solid curves represent quadratic functions
used in the chiral extrapolation.
Two dashed lines indicate the $\pi\Sigma$ and the $\bar K N$ thresholds
in the presence of the relative momentum $p=\frac{2\pi}{L}$.
Two solid lines on the vertical axes show
the experimentally observed masses of $\Lambda(1115)$ and $\Lambda(1600)$.
}
\end{figure}

\begin{figure}[hbt]
\includegraphics[scale=0.35]{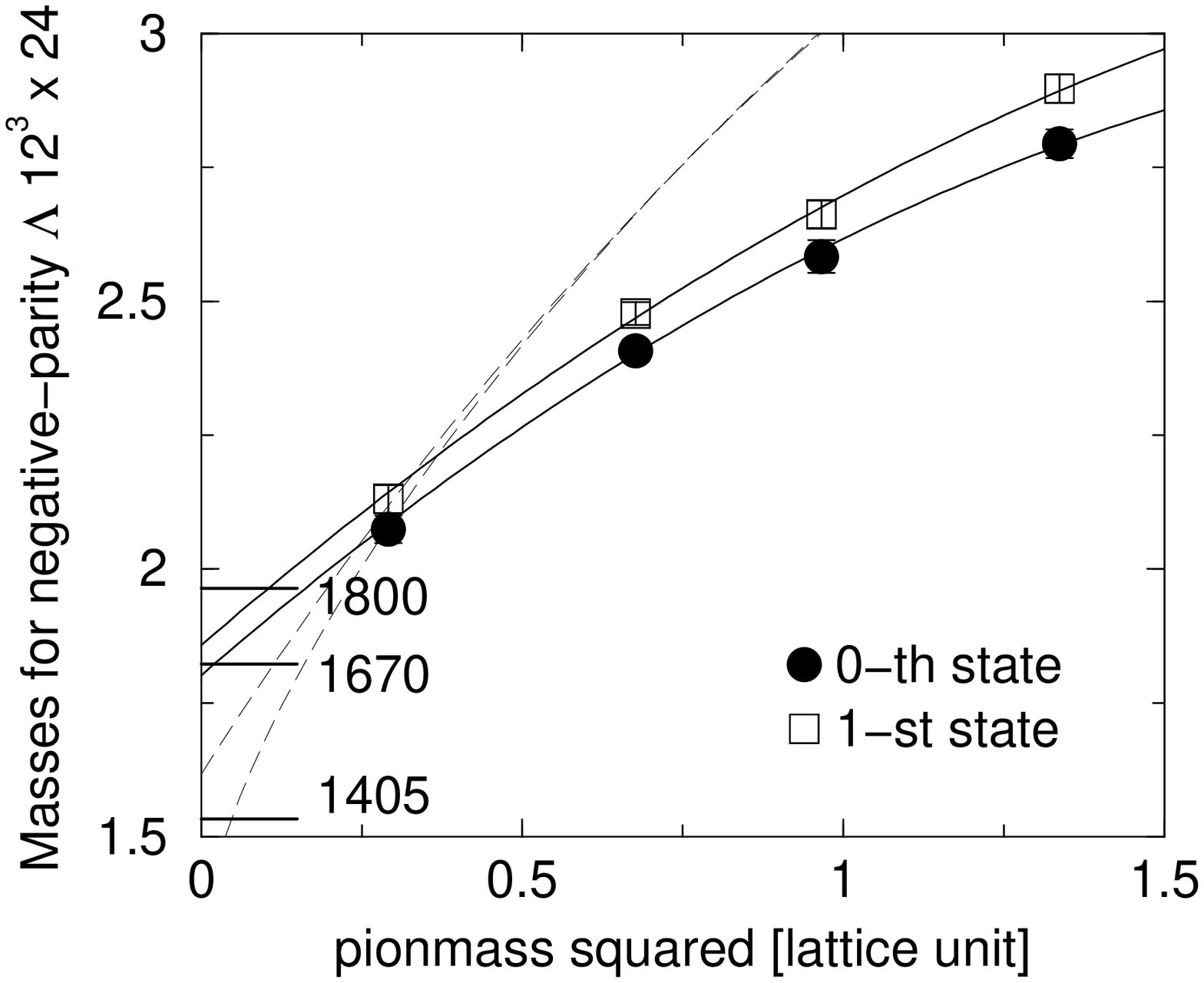}
\includegraphics[scale=0.35]{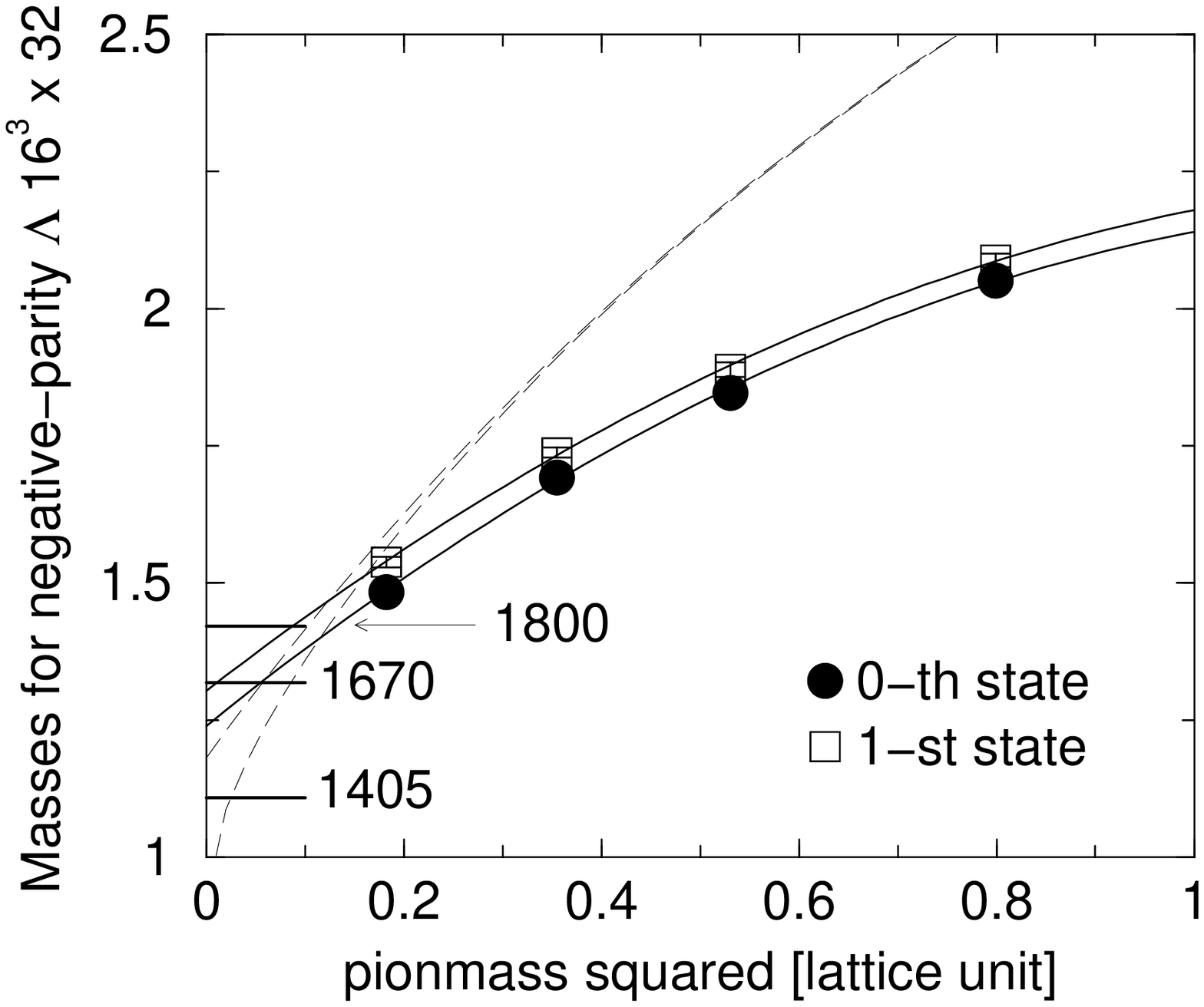}
\includegraphics[scale=0.35]{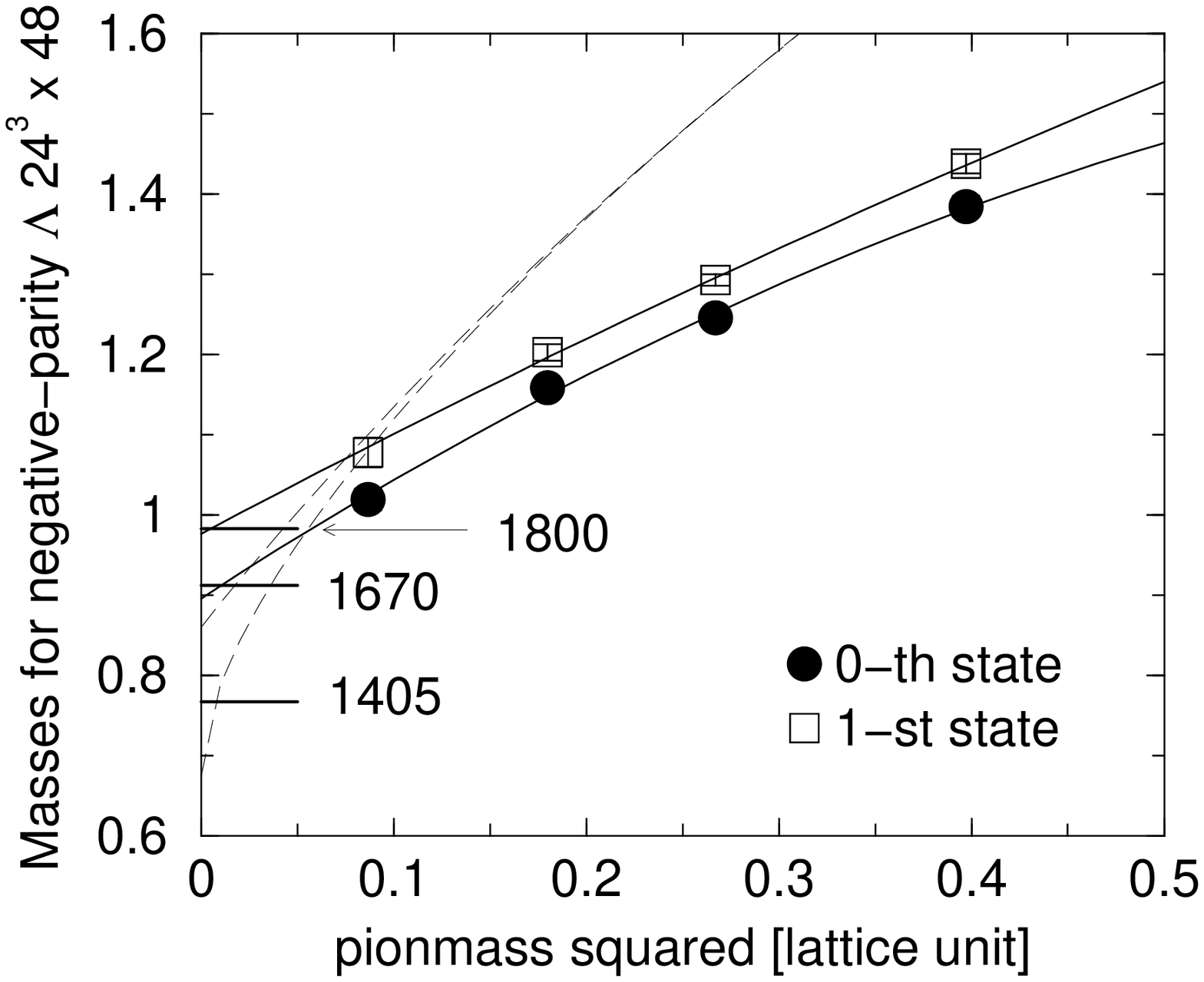}
\caption{\label{negmass}
Masses of the lowest two negative-parity $\Lambda$ states
as functions
of the squared pion mass. The filled circles (open squares) denote
the masses of the lowest (1st-excited) state.
Two solid curves represent quadratic functions
used in chiral extrapolation.
Two dashed lines indicate the $\pi\Sigma$ and $\bar K N$ thresholds.
Two solid lines on the vertical axes show
the experimentally observed masses of $\Lambda(1405)$, $\Lambda(1670)$ and $\Lambda(1800)$.
}
\end{figure}

In Fig.~\ref{effmassexample}, as typical results, 
effective masses are plotted
for the ground and the 1st-excited states obtained from 
positive- and negative-parity projected data with 
$\kappa_{\rm sea}, \kappa_{\rm val} = 0.1367$ at $\beta = 2.10$.
We can find $t$-windows in which the effective masses exhibit plateaus.
We extract eigen-energies by fitting lattice QCD data
in the $t$-range where both of the ground- and 1st-excited states
exhibit plateaus.
In Table~\ref{hadronicmass}, we list all the hadronic masses
extracted in this study.
Figs.~\ref{posmass} and \ref{negmass} show
the eigen-energies in the positive- and negative-parity channels,
respectively,
plotted as functions of the squared pion mass $m_\pi^2$.
Filled circles denote the energies of the ground states,
and open squares those of 1st-excited states.
The solid curves represent quadratic fits as a function of squared pion
mass $m_\pi^2$.
Two solid lines at the vertical axes
indicate the masses of $\Lambda(1115)$ and $\Lambda(1600)$
in Fig.~\ref{posmass} (positive-parity states),
and $\Lambda(1405)$, $\Lambda(1670)$ and $\Lambda(1800)$
in Fig.~\ref{negmass} (negative-parity states).

The obtained masses of  the positive-parity ground state
agree very well with the mass of the ground-state $\Lambda(1115)$
at all three $\beta$'s.
On the other hand, the 1st-excited state in this channel
lies much higher than $\Lambda(1600)$,
which is the 1st excited state experimentally observed so far.
The same tendency was reported in Ref.~\cite{Burch:2006cc}, and 
the situation is similar to the case of  
the Roper resonance, 
which is the non-strange SU(3) partner of $\Lambda (1600)$~\cite{Sasaki:2001nf}.

In the negative-parity channel (Fig.\ref{negmass}),
the ground- and the 1st-excited states always have close energies
at all the $\beta$'s and $\kappa$'s.
The eigen-energies have similar quark-mass dependences, and
the mass splittings are almost quark mass independent.
The chirally extrapolated values both lie around 
the mass of $\Lambda(1670)$ rather than $\Lambda(1405)$.
Similarly to the previous studies,
we do not reproduce the mass of $\Lambda(1405)$ in our calculation.
While in quenched simulations in Refs.\cite{Melnitchouk:2002eg,Nemoto:2003ft}
such failure was regarded as an evidence
of possible meson-baryon molecule components in $\Lambda(1405)$,
our present {\it unquenched}
simulation contains effects of dynamical quarks
and thus should incorporate meson-baryon molecular states.
We will come back to this point later in section V.


\begin{table}[h]
\begin{tabular}{rrrrrrrrr}
\hline
$\beta$ & $\kappa$ & 
$m_\pi$  & $m_K$   & $m_N$   & $m_\Sigma$
\\ \hline\hline

1.80 & 0.1409 & 1.156(1) & 1.087(01) & 2.260( 5) & 2.201(05) \\

     & 0.1430 & 0.983(1) & 0.976(01) & 2.024(07) & 2.019(06) \\

     & 0.1445 & 0.822(1) & 0.872(01) & 1.813(09) & 1.844(07) \\

     & 0.1464 & 0.531(2) & 0.672(02) & 1.405(11) & 1.530(09) \\\hline

     & {\rm C.L.} & --- & 0.291(30) & 1.078(41) & 1.269(17) \\\hline\hline

1.95 & 0.1375 & 0.894(1) & 0.834(01) & 1.707(06) & 1.468(06) \\

     & 0.1390 & 0.729(1) & 0.725(01) & 1.475(05) & 1.464(06) \\

     & 0.1400 & 0.596(1) & 0.635(01) & 1.289(05) & 1.318(05) \\

     & 0.1410 & 0.427(1) & 0.533(01) & 1.051(08) & 1.134(07) \\\hline

     & {\rm C.L.} & --- & 0.455(16) & 0.777(27) & 0.919(18) \\\hline\hline

2.10 & 0.1357 & 0.630(1) & 0.565(1) & 1.194(05) & 1.131(06) \\

     & 0.1367 & 0.517(1) & 0.491(1) & 1.016(05) & 0.990(05) \\

     & 0.1374 & 0.424(1) & 0.435(1) & 0.902(05) & 0.907(05) \\

     & 0.1382 & 0.295(1) & 0.363(1) & 0.734(06) & 0.784(05) \\\hline

     & {\rm C.L.} & --- & 0.312(9) & 0.579(32) & 0.674(26) \\\hline\hline

\end{tabular}

\vspace{.5cm}
\begin{tabular}{rrrrrr}
\hline
$\beta$ & $\kappa$ & 
$E_0^+$  & $E_1^+$ & $E_0^-$ & $E_1^-$ 
\\ \hline\hline

1.80 & 0.1409 & 
2.201(03) & 3.53(34) & 2.794(27) & 2.897(27) \\

     & 0.1430 & 
2.013(04) & 3.53(40) & 2.583(31) & 2.662(26) \\

     & 0.1445 & 
1.836(04) & 2.99(18) & 2.407(20) & 2.477(21) \\

     & 0.1464 & 
1.533(07) & 2.22(19) & 2.074(27) & 2.131(26) \\\hline

     & {\rm C.L.} &
1.284(14) & 1.40(19) & 1.801(31) & 1.858(41) \\\hline\hline

1.95 & 0.1375 & 
1.640(3) & 2.465(28) & 2.051(13) & 2.088(12) \\

     & 0.1390 & 
1.457(3) & 2.292(72) & 1.847(13) & 1.889(13) \\

     & 0.1400 & 
1.309(2) & 2.094(18) & 1.692(09) & 1.737(09) \\

     & 0.1410 & 
1.134(3) & 1.899(19) & 1.483(12) & 1.538(10) \\\hline

     & {\rm C.L.} &
0.925(6) & 1.656(22) & 1.239(30) & 1.303(23) \\\hline\hline

2.10 & 0.1357 & 
1.133(09) & 1.677(28) & 1.384(13) & 1.437(12) \\

     & 0.1367 & 
0.994(05) & 1.526(15) & 1.246(09) & 1.293(07) \\

     & 0.1374 & 
0.919(05) & 1.469(18) & 1.158(10) & 1.203(10) \\

     & 0.1382 & 
0.778(06) & 1.325(15) & 1.019(14) & 1.078(18) \\\hline

     & {\rm C.L.} &
0.653(52) & 1.214(69) & 0.896(38) & 0.977(31) \\\hline\hline

\end{tabular}
\caption{\label{hadronicmass}
Hadronic masses (in lattice units) obtained for four different $\kappa$'s at each $\beta$. 
The $m_\pi$, $m_K$, $m_N$, and $m_\Sigma$
denote the masses of pion, kaon, nucleon, and $\Sigma$ baryon,  respectively.
$E_i^\pm$ represents
the eigen-energy of the $i$-th state 
in positive ($+$) or negative ($-$) parity channel with $I=0$ and $S=-1$.
``C.L.'' in each block shows the results extrapolated to the chiral limit
with quadratic functions.
}
\end{table}

\begin{table}[h]
\begin{tabular}{rrrrrr}
\hline
$\beta$ & $\kappa$ & 
$g_0^+\times 10^2$  &$g_1^+\times 10^2$  &$g_0^-\times 10^2$  &$g_1^-\times 10^2$ \\ \hline\hline

1.80 & 0.1409 &
 4.91(303)  &  0.30(01) & $-$10.7(036) &   6.46(291) \\ 

     & 0.1430 &
 0.12(013)  &  0.02(00) &  $-$0.64(035) &   0.40(027) \\ 

     & 0.1445 &
$-$3.82(195)  & $-$0.33(01) &  15.2(023) & $-$10.5(018) \\ 

     & 0.1464 &
 0.10(126) & $-$1.39(06) &  35.7(077) & $-$21.9(063) \\\hline

     & {\rm C.L.} &
$-$8.36(800) & $-$1.66(27) &  65.3(116) & $-$43.1(113) \\\hline\hline

1.95 & 0.1375 &
 7.22(066) &  0.46(01) & $-$33.4(145) &  22.2(109)\\ 

     & 0.1390 &
 0.58(029) &  0.11(00) &  $-$2.99(141) &   4.25(167)\\ 

     & 0.1400 &
$-$2.46(022) & $-$0.32(00) &  11.4(013) &  $-$8.83(120)\\ 

     & 0.1410 &
$-$7.02(081) & $-$1.14(01) &  26.3(031) & $-$15.2(025)\\\hline

     & {\rm C.L.} &
$-$9.12(323) & $-$1.81(24) &  39.9(049) & $-$19.8(072)\\ \hline\hline

2.10 & 0.1357 &
 10.07(128) &  0.78(03) & $-$22.3(038) & 12.4(036)\\ 

     & 0.1367 &
  5.17(031) &  0.37(01) & $-$17.9(061) & 8.21(367)\\ 

     & 0.1374 &
 $-$0.77(010) & $-$0.09(00) & 2.19(051) & $-$1.26(044)\\ 

     & 0.1382 &
 $-$5.44(112) & $-$1.11(04) & 19.8(036) & $-$15.2(040)\\ \hline

     & {\rm C.L.} &
 $-$14.4(052) & $-$1.81(47) & 42.5(103) & $-$33.3(017)\\ \hline\hline

\end{tabular}
\caption{\label{hadronicmass2}
The ratios $g_i^\pm$ (multiplied by $10^2$) defined
in Eqs.(\ref{gminus0})-(\ref{gplus1}).
``C.L.'' shows the values extrapolated to the chiral limit
using quadratic functions.
}
\end{table}

\subsection{lattice-spacing dependences}

\begin{figure}[hbt]
\includegraphics[scale=0.35]{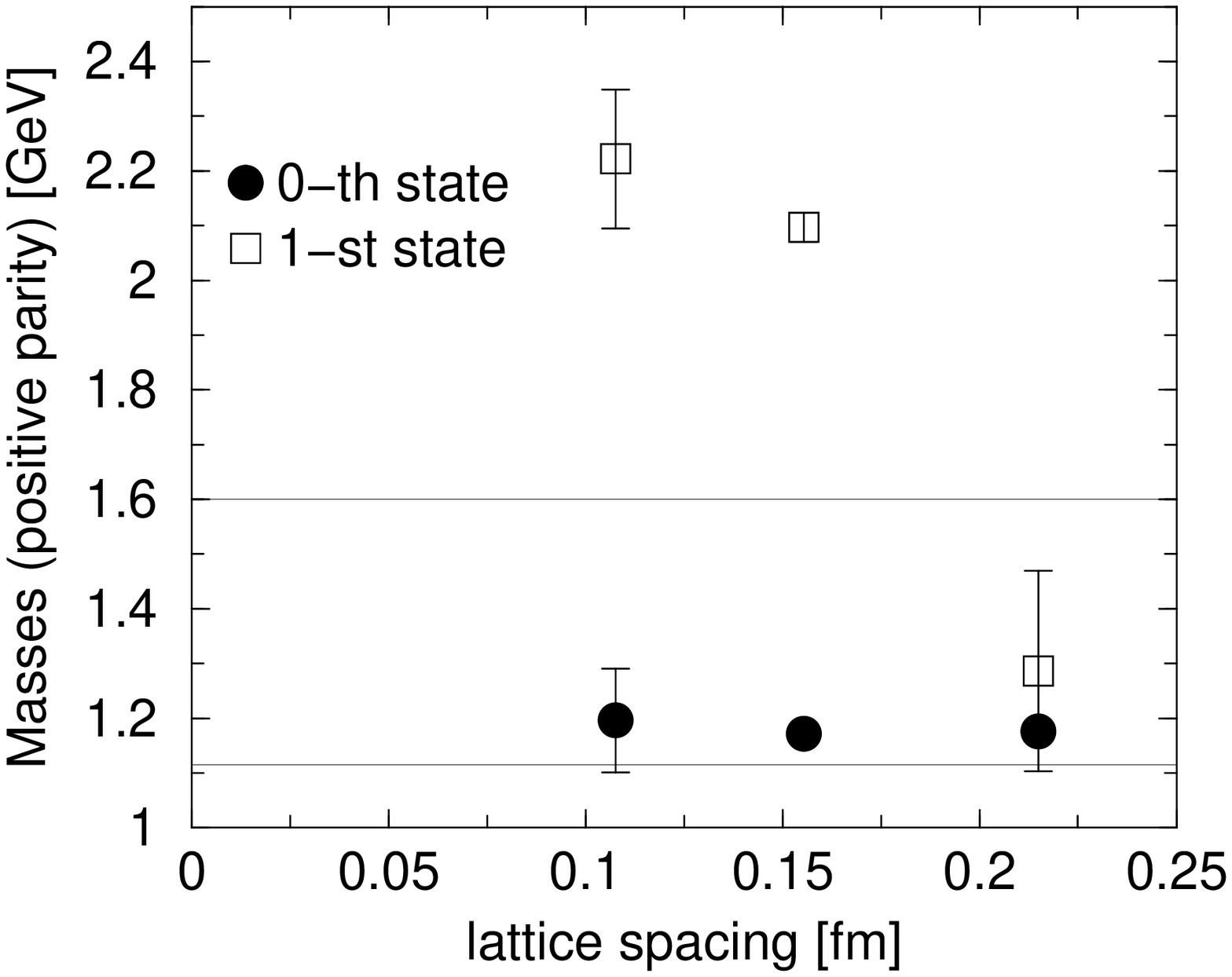}
\includegraphics[scale=0.35]{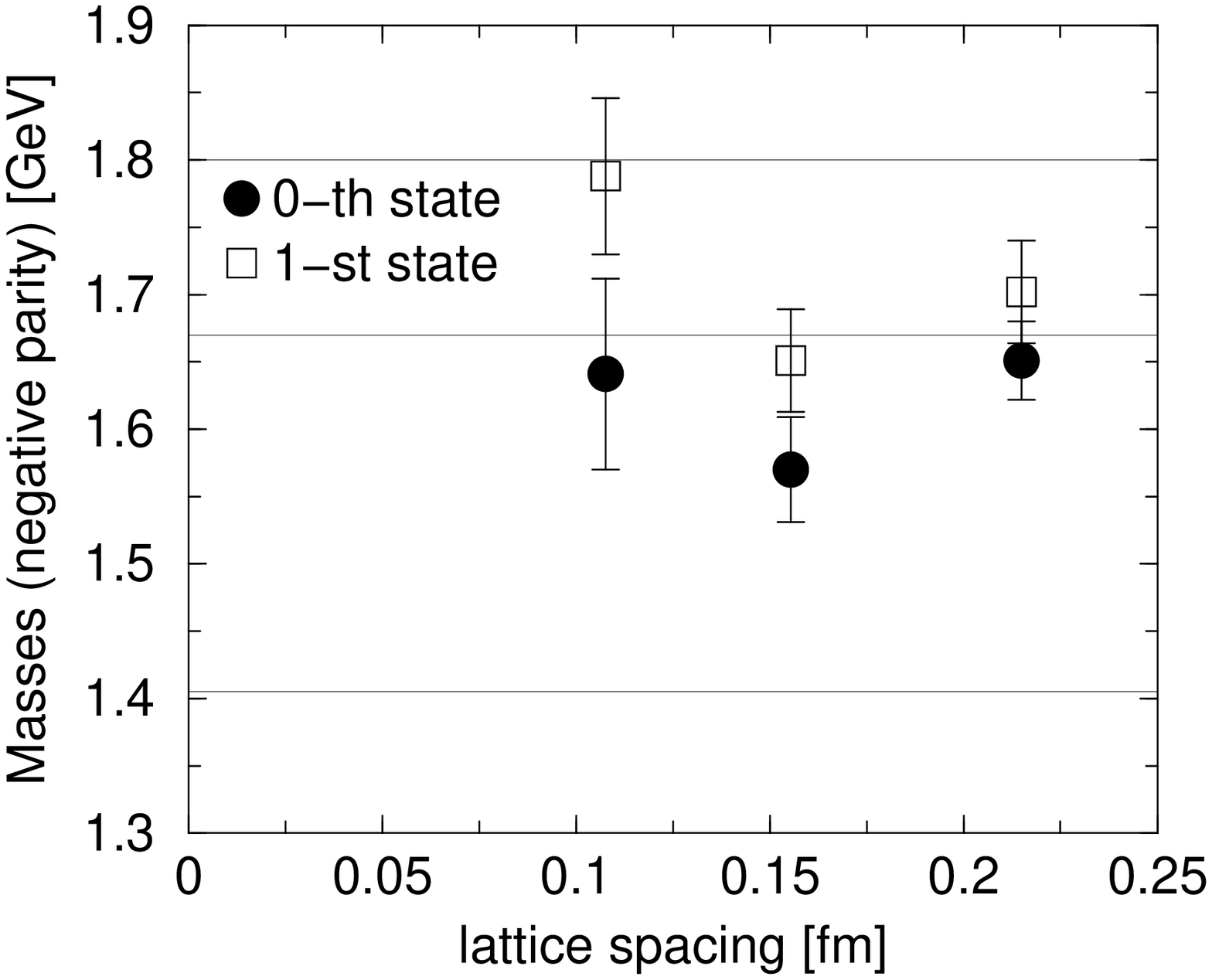}
\caption{\label{massvsa}
Masses of the ground and the 1st-excited
states in the chiral limit 
for the positive (upper) and negative (lower) parity $\Lambda$ states
plotted as functions of the lattice spacing $a$.
The horizontal lines
indicate the masses of the experimentally observed $\Lambda$ states.
}
\end{figure}

Fig. \ref{massvsa} shows the chirally extrapolated masses 
of the lowest two states in each parity
v.s. the lattice spacing $a$.
Except for the 1st-excited positive-parity state at the largest $a$,
all the data show little lattice-cutoff dependence.
The horizontal lines indicate the masses of the experimentally 
observed $\Lambda$ states.
The positive-parity ground state
agrees well with the mass of $\Lambda(1115)$, while
the lowest two negative-parity states lie around $\Lambda(1670)$.

\subsection{possible contaminations by scattering states}

Since our calculations contain
dynamical u and d quarks,
scattering states could come into spectra.
$N\bar K$ and $\Sigma\pi$ thresholds
are drawn in Fig.~\ref{posmass} and Fig.\ref{negmass} as dashed lines.
The energy of the 1st-excited state 
at the lightest u- and d-quark masses
could be contaminated by such scattering states.
Several strategies have been adopted to discriminate
resonances from scattering states:
study of volume dependence of the masses, and the spectral weights
or boundary condition dependence of the eigen-energies.
They were tested and applied to the pentaquark baryon
~\cite{Sasaki:2006jn,Mathur:2003zf,Csikor:2003ng,
Mathur:2004jr,Ishii:2004qe,Takahashi:2005uk}.

There are two meson-baryon channels relevant in the present calculation,
$\pi\Sigma$ and $\bar K N$.
In terms of the valence quark contents,
we have 5 different thresholds
(2 for $\bar K N$, 3 for $\pi\Sigma$).
In this paper, 
in order to distinguish resonance states from all these scattering states,
we impose the following boundary condition on the quark fields,
\begin{equation}
\psi(x+L)
=
e^
{\frac23\pi i}
\psi(x).
\end{equation}
Under such boundary condition for quark fields,
a hadronic state $\phi_{3k+n}(x)$ which contains $3k+n$ valence quarks obeys
\begin{equation}
\phi_{3k+n}(x+L)
=
e^
{\frac23n\pi i}
\phi_{3k+n}(x),
\end{equation}
and can have spatial momenta,
\begin{eqnarray}
p_{\rm lat}
=
\frac{2\pi}{L}m+\frac{2n\pi}{3L}
\ \ 
(m \in {\rm Z})
\end{eqnarray}

As a result,
only states which consist of 3k valence quarks
can be zero-momentum states.
Since other quark combinations 
inevitably have non-vanishing spatial momenta, their energies are raised up.
As long as we employ three-quark operators for baryon creation/annihilation,
hadronic states appearing in scattering states should contain
one or two valence quark(s), 
since sea-quark pairs themselves cannot carry flavors.

\begin{figure}[hbt]
\includegraphics[scale=0.35]{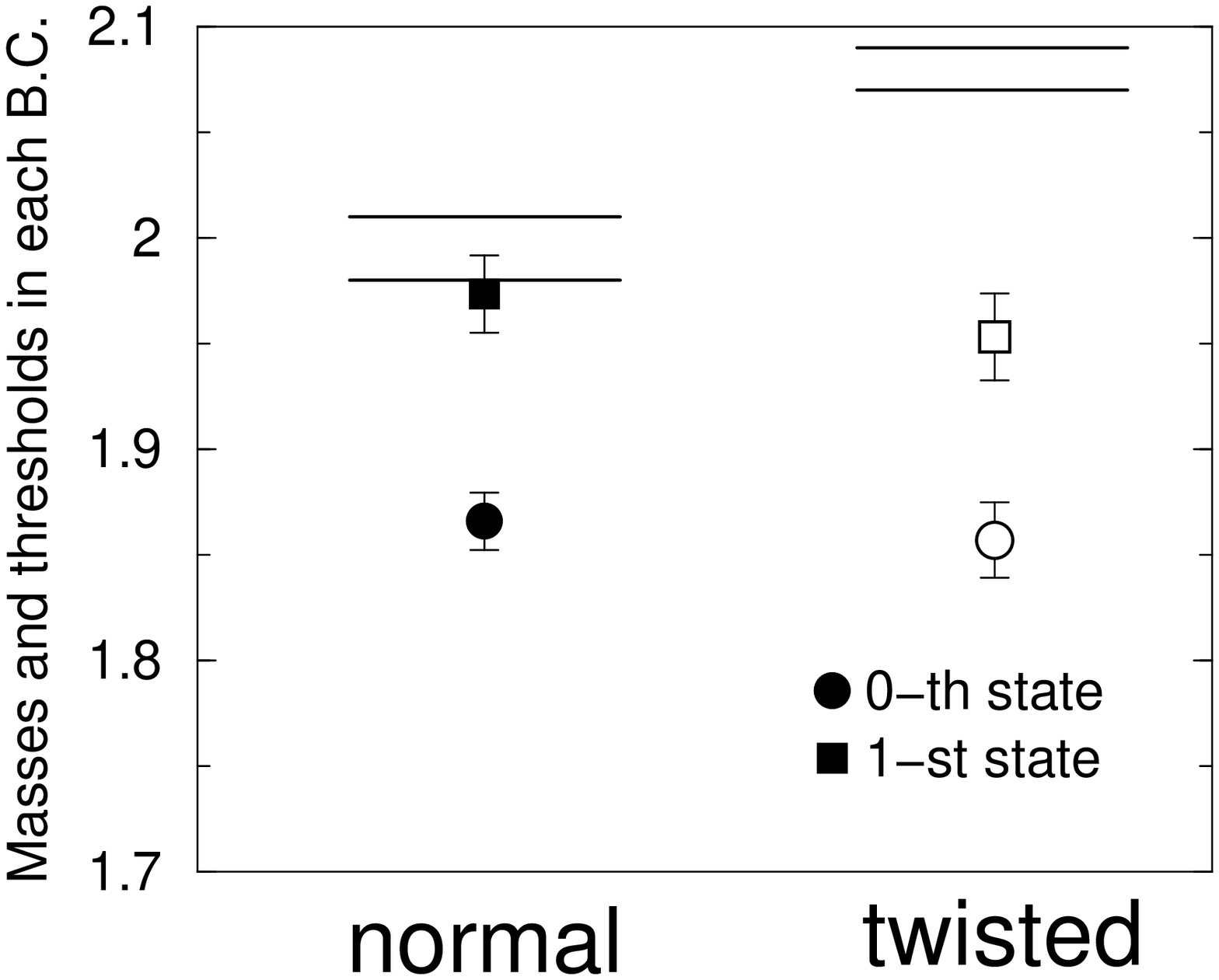}
\caption{\label{compare}
Lowest two eigen-energies 
in the $(J^P,S) = (1/2^-,-1)$ channel
at $\beta =2.10$ and $\kappa =0.1382$ 
under the normal (periodic) and the twisted boundary conditions.
The circles (squares) are for the ground (1st-excited) states.
The lower and upper solid lines respectively represent
the threshold energies of $\Sigma\pi$ and $N\bar K$ states
evaluated with normal/twisted boundary condition.}
\end{figure}

Now we turn back to our full-QCD calculations.
We impose the following boundary condition to all the valence
quark fields ($q=u,\ d,\ s$);
\begin{equation}
q(x+L)
=
e^
{\frac23\pi i}
q(x).
\end{equation}

We plot in Fig.~\ref{compare} the eigen-energies under 
the periodic and the twisted boundary conditions,
which are obtained at $\beta =2.10$ and $\kappa =0.1382$.
The open and filled circles (squares) are for the ground (1st-excited) states.
The lower and upper solid lines respectively represent
the threshold energies of $\Sigma\pi$ and $N\bar K$ states
with normal/twisted boundary condition.
The threshold energies are raised up
in the case of twisted boundary condition,
whereas the lowest two eigen-energies remain unchanged.
Thus we conclude that the observed states are insensitive to boundary conditions,
and contaminations from meson-baryon scattering states are small.
Namely the results indicate that these states are regarded as compact resonance states.

Similar results are obtained also for the positive-parity states.
Their energies are insensitive to the boundary conditions.
(The energies of the positive-parity 1st-excited states at the smallest $\kappa$
are higher than 
p-wave thresholds (with relative momenta $p = \frac{2\pi}{L}$)
of $\Sigma\pi$ or $N\bar K$ states.)

\section{Flavor structures}

The chiral unitary approach~\cite{Jido:2003cb} has suggested
that $\Lambda (1405)$ is not a single pole
but a superposition of two independent resonance poles.
The structure of $\Lambda (1405)$
is now attracting much interest,
and desired to be clarified in a model independent manner.
We investigate the flavor structure
of the ground and the 1st-excited states
obtained from the cross correlators of two operators.

In order to clarify the flavor structure of the low-lying states,
we evaluate the overlaps of the obtained states with the source and
sink operators.
We define the magnitudes of the singlet ($I={\bf 1}$) and the octet
($I={\bf 8}$)
components in the $i$-th state ($i=0$ for the ground state and $i=1$ for the
1st excited state) by 
\begin{equation}
\langle i | \eta_I^\dagger | {\rm vac}\rangle = (C_{\rm})_{iI}.
\end{equation}
Actually, we can only obtain the ratio between
\begin{equation}
\langle i | \eta_{\bf 1}^\dagger | {\rm vac}\rangle = (C_{\rm})_{i{\bf 1}}
\end{equation}
and
\begin{equation}
\langle i | \eta_{\bf 8}^\dagger | {\rm vac}\rangle = (C_{\rm})_{i{\bf 8}},
\end{equation}
since the overall factors in $C_{\rm src}$ cannot be determined
in our setups.
We evaluate $g_0$ and $g_1$ defined as
\begin{eqnarray}
g_0^-
&\equiv&
C_{0{\bf 8}}/C_{0{\bf 1}} 
\label{gminus0}
\\
g_1^-
&\equiv&
C_{1{\bf 1}}/C_{1{\bf 8}},
\label{gminus1}
\end{eqnarray}
Both $g_0^-$ and $g_1^-$ vanish when the SU(3)$_F$ symmetry is exact,
showing that the ground (1st excited) state is purely flavor singlet (octet) 
in the limit.
For the positive-parity states, we similarly define
\begin{eqnarray}
g_0^+
&\equiv&
C_{0{\bf 1}}/C_{0{\bf 8}} 
\label{gplus0}
\\
g_1^+
&\equiv&
C_{1{\bf 8}}/C_{1{\bf 1}}.
\label{gplus1}
\end{eqnarray}
In this case, we exchange the denominator and the numerator
since the ground state is flavor octet and
the 1st excited state is flavor singlet in the SU(3)$_F$ limit.

\begin{figure}[hbt]
\includegraphics[scale=0.35]{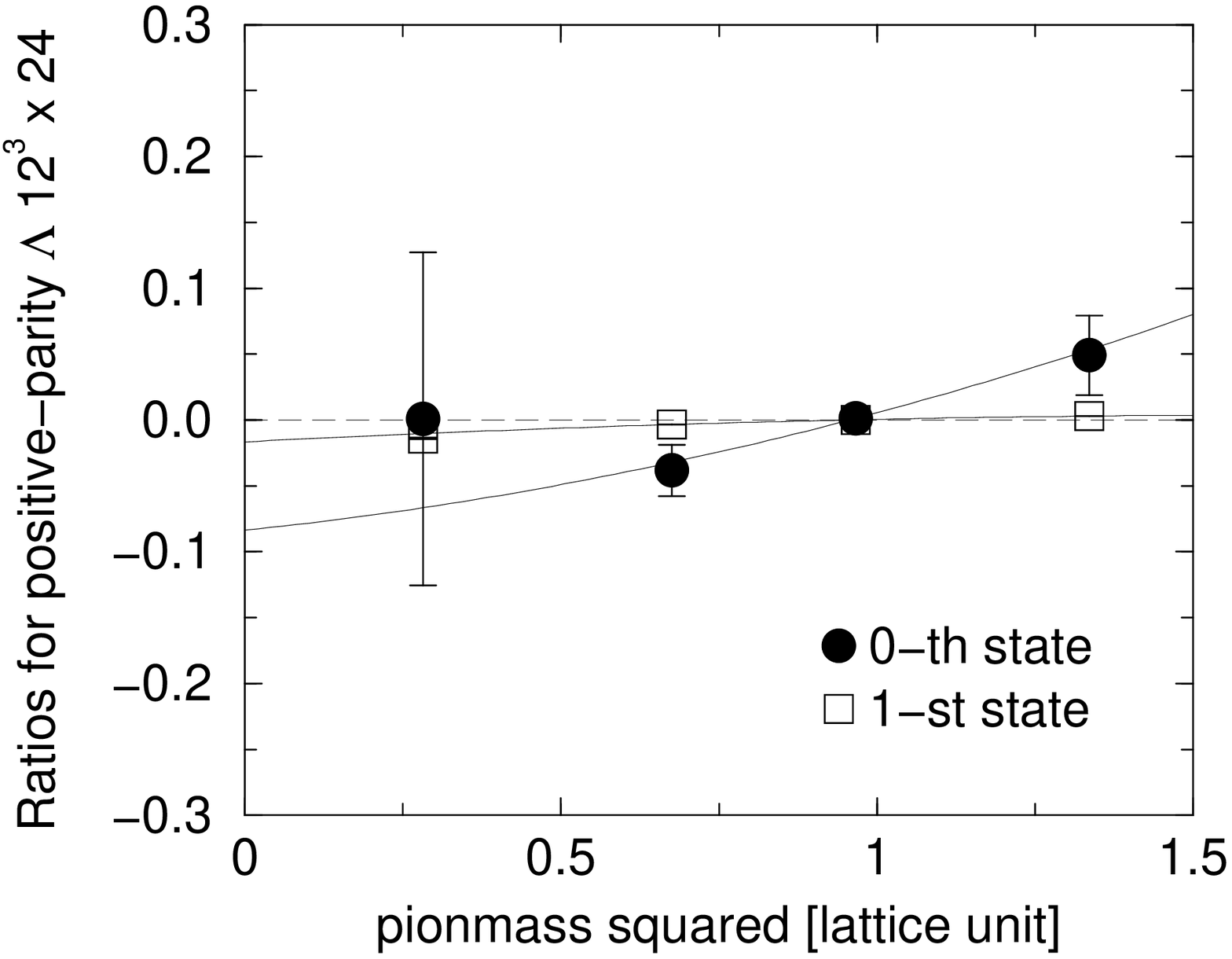}
\includegraphics[scale=0.35]{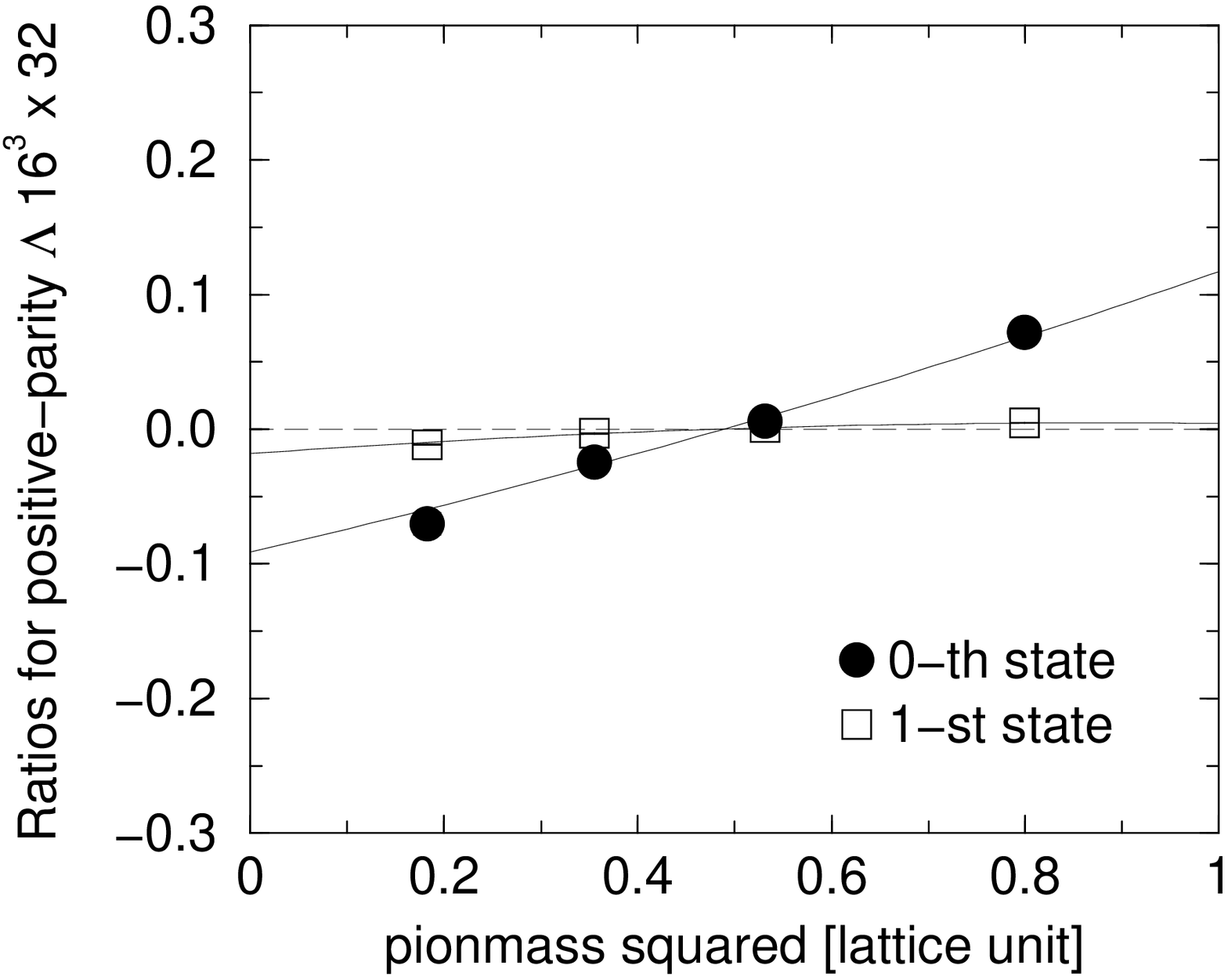}
\includegraphics[scale=0.35]{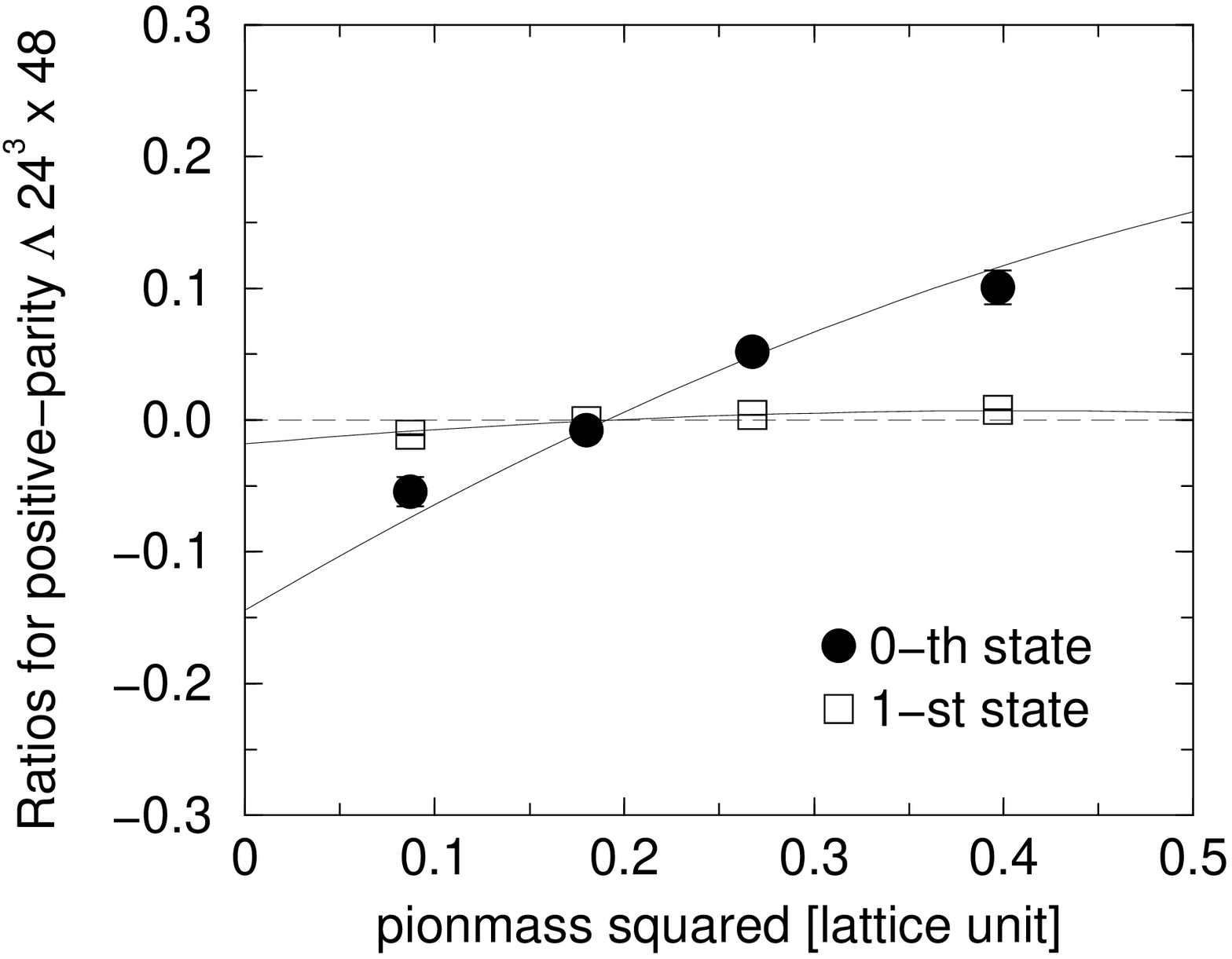}
\caption{\label{gpos}
$g_i^+$ are plotted as a function of $m_\pi^2$.
Solid lines denote quadratic functions used in chiral extrapolation.
}
\end{figure}
\begin{figure}[hbt]
\includegraphics[scale=0.35]{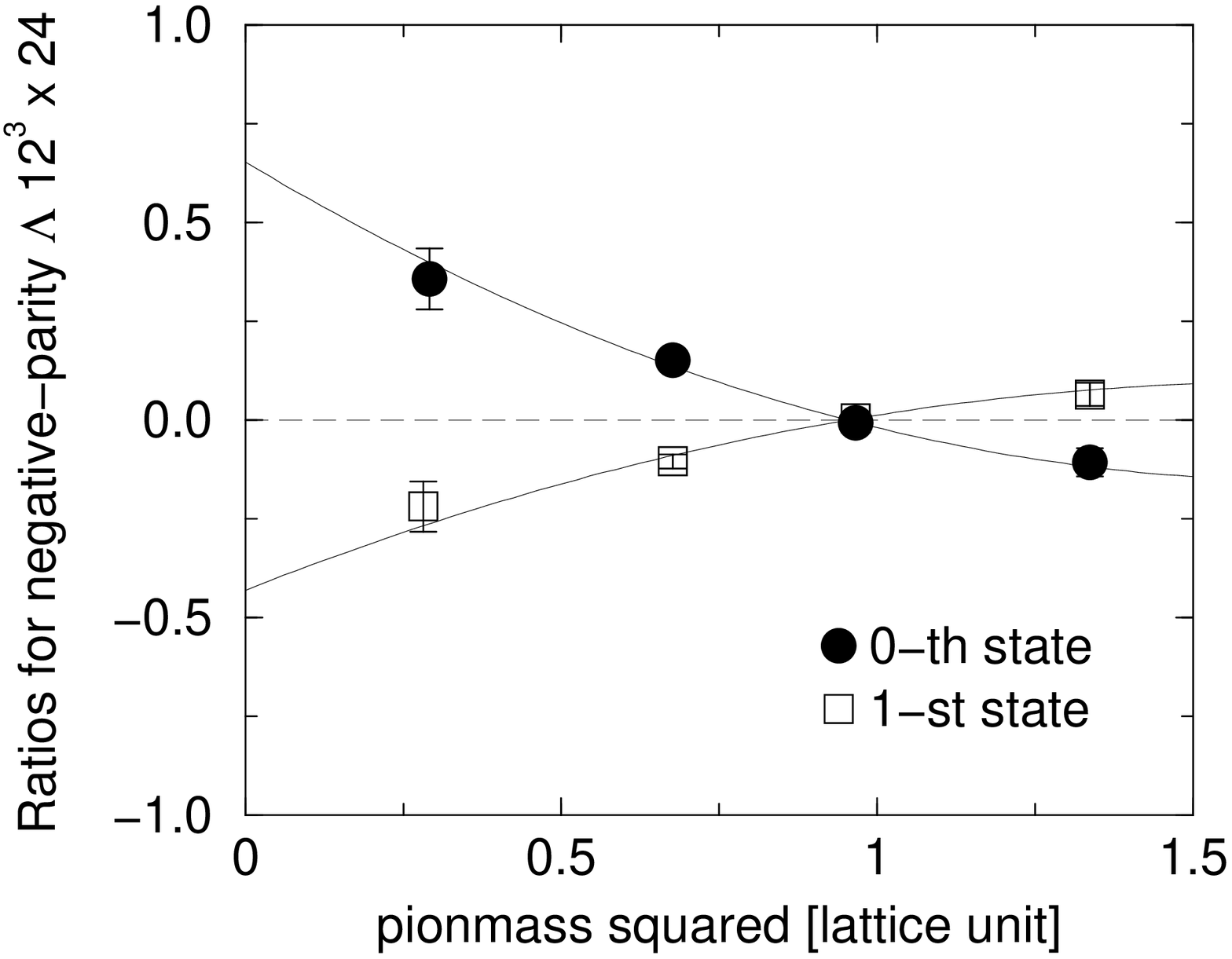}
\includegraphics[scale=0.35]{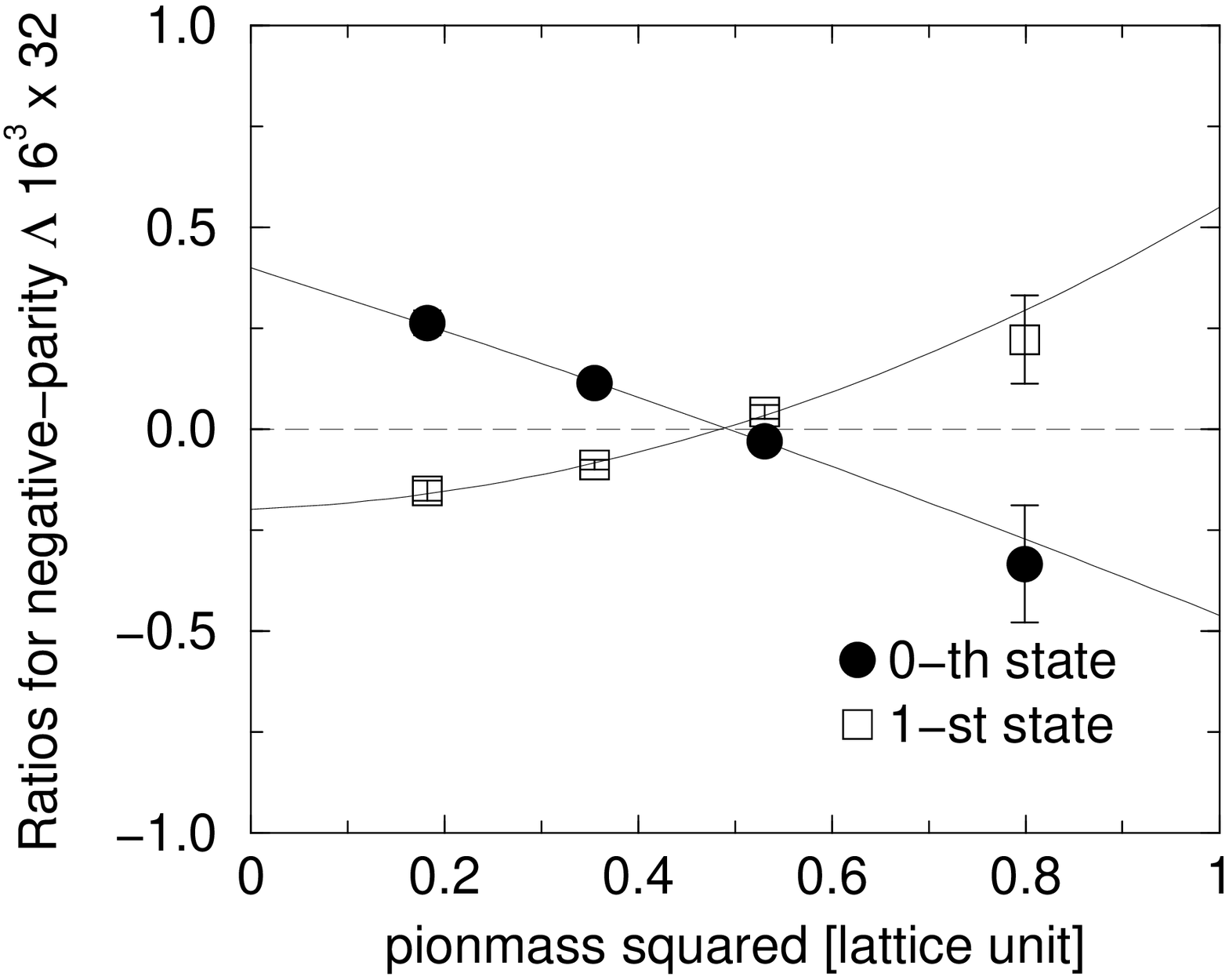}
\includegraphics[scale=0.35]{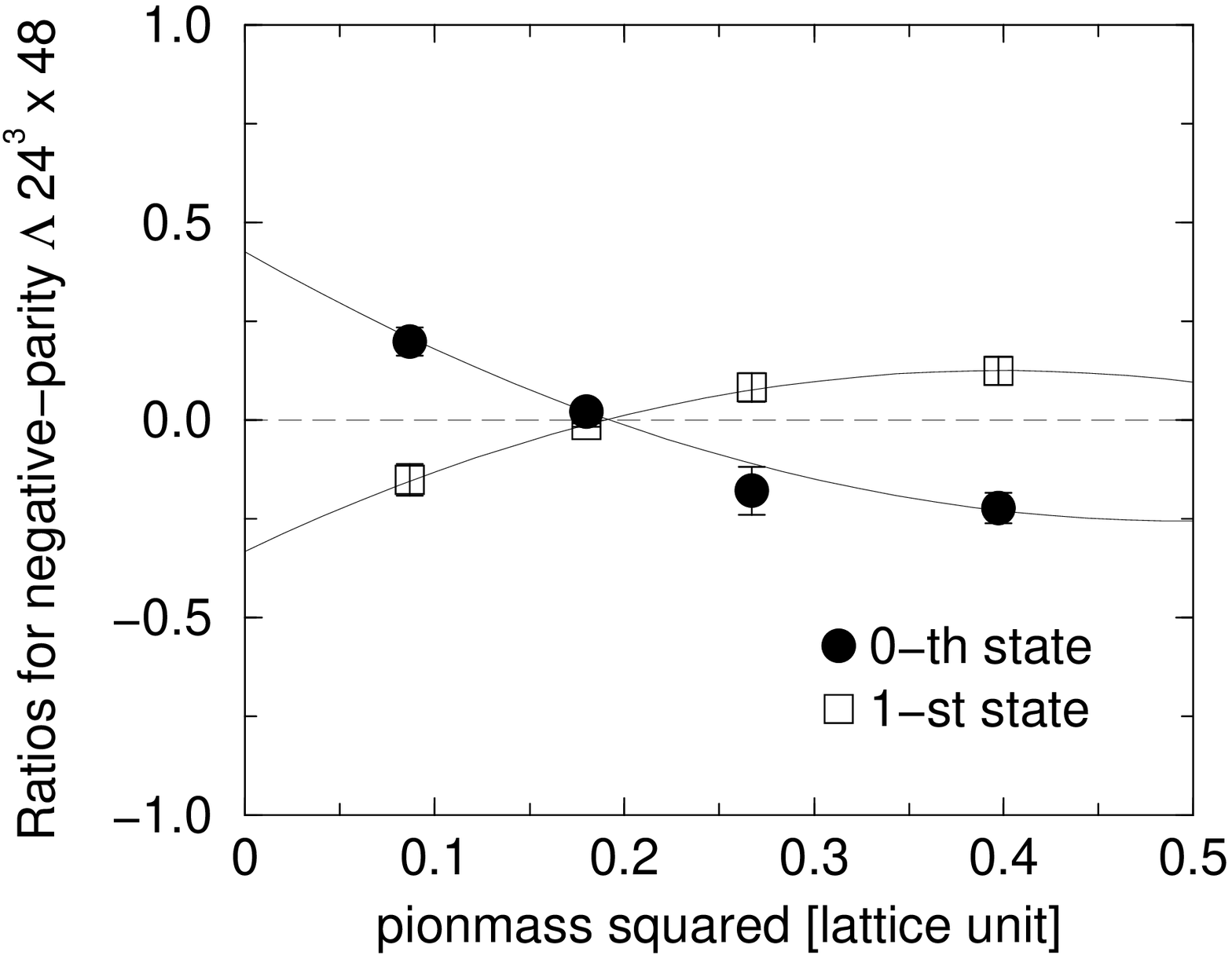}
\caption{\label{gneg}
$g_i^-$ are plotted as a function of $m_\pi^2$.
Solid lines denote quadratic functions used in chiral extrapolation.
}
\end{figure}

We show $g_0^\pm$ and $g_1^\pm$
as a function of pion-mass squared
for each lattice cutoff in Fig.~\ref{gpos} and Fig.~\ref{gneg} .
As is expected from the symmetry, such mixing coefficients
cross zero at the flavor symmetric limit
($\kappa_u=\kappa_d=\kappa_s$).
The data show smooth quark-mass dependence
toward the chiral limit in all the lattice-cut-off cases
and 
the dependences are almost lattice-cut-off independent.
We find that
the magnitude of operator mixing
gets larger for larger flavor-symmetry breaking.
It is interesting that
the 1st-excited state in positive-parity channel 
is dominated by the flavor-singlet component
showing almost no contaminations of octet components.

We perform quadratic fits
of these mixing coefficients
as
\begin{eqnarray}
g_0(m_\pi^2)
=
a_{2,0} m_\pi^2
+
a_{1,0} m_\pi
+
a_{0,0} \\
g_1(m_\pi^2)
=
a_{2,1} m_\pi^2
+
a_{1,1} m_\pi
+
a_{0,1}
\end{eqnarray}
and list the extrapolated values in Table~\ref{hadronicmass2}.
Even in the chiral limit, 
$g_i\ (i=1,2)$ in negative-parity channel remain less than 0.6,
which indicates that
the ground (1st-excited) state
is dominated by the singlet (octet) component
in the present quark-mass range.

On the other hand, the mixings of the singlet and octet components
are generally weaker in the positive parity $\Lambda$'s.
It is natural because the splitting of the two states are large in the
SU(3)$_F$.
Nevertheless, it is interesting to see that the ground state, 
corresponding to $\Lambda(1115)$, contains significant (about 10\%) 
flavor singlet component in the chiral limit, which was not
expected from the simple quark model picture.

In practice, we need to interpret such overlap coefficients
into physical amplitudes.
In the Euclidean-time range we consider
\begin{eqnarray}
\eta_{\bf 1} | {\rm vac} \rangle
=
c_0|0\rangle + g_1c_1|1\rangle
\\
\eta_{\bf 8} | {\rm vac} \rangle
=
g_0c_0|0\rangle + c_1|1\rangle
\end{eqnarray}
are satisfied. We here assume that
each state can be expanded
in terms of the singlet and octet basis states as
\begin{eqnarray}
&&|0\rangle
=
M_{01} |\Lambda_1\rangle
+
M_{08} |\Lambda_8\rangle \\
&&|1\rangle
=
M_{11} |\Lambda_1\rangle
+
M_{18} |\Lambda_8\rangle.
\end{eqnarray}
We note that $|\Lambda\rangle$'s
in this expression
are no longer energy eigenstates in QCD,
and they are restricted to elements
that interpolating operators have overlaps with;
$|\Lambda\rangle\propto\eta^\dagger|{\rm vac}\rangle$.
(Actually we can introduce other singlet and octet bases, for example,
whose spatial distributions are different.
If we adopt more interpolating fields
such as those with spatial derivatives or with exotic spinor structures,
we can introduce much more basis states in this argument.)

\begin{figure}[h]
\includegraphics[scale=1]{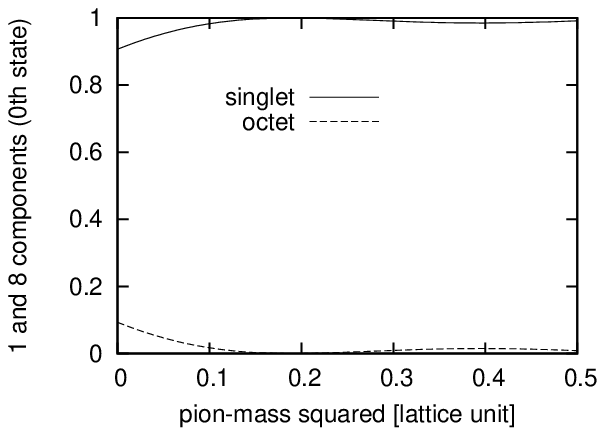}
\includegraphics[scale=1]{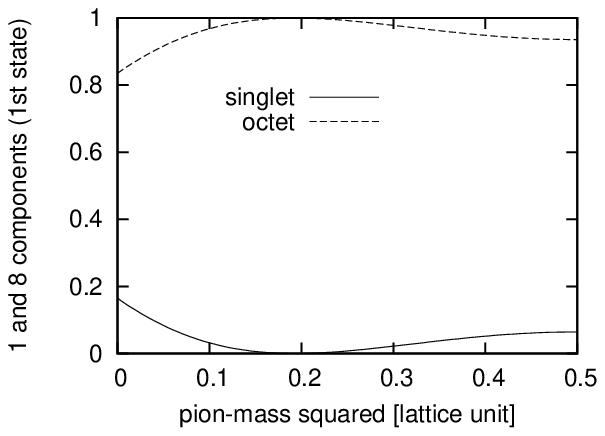}
\caption{\label{FcompoN}
$M_{ij}^2$ 
obtained at $\beta = 2.1$
are plotted as a function of squared pion mass $m_\pi^2$.
$M_{ij}$ are normalized so that $M_{i1}^2+M_{i8}^2=1$.
}
\end{figure}

We here extract $M_{ij}$'s, using the conditions, 
$\langle \Lambda_8 | \eta_{\bf 1} | {\rm vac} \rangle = 0$
and
$\langle \Lambda_1 | \eta_{\bf 8} | {\rm vac} \rangle = 0$,
giving
\begin{eqnarray}
&&\frac{M_{08}}{M_{18}}
=
-g_1R,
\\
&&\frac{M_{11}}{M_{01}}
=
-g_0R^{-1}
\end{eqnarray}
with $R = c_1/c_0$.
$R^2$ can be estimated from the 2-point correlators,
since optimized operators satisfy
\begin{eqnarray}
&&(\eta_{\bf 1}-g_1\eta_{\bf 8})| {\rm vac}\rangle
=
c_0(1-g_0g_1)|0\rangle
\\
&&(\eta_{\bf 8}-g_0\eta_{\bf 1})| {\rm vac}\rangle
=
c_1(1-g_0g_1)|1\rangle.
\end{eqnarray}
Using the normalization relations,
\begin{eqnarray}
&&
M_{01}^2+M_{08}^2 = 1
\\
&&
M_{11}^2+M_{18}^2 = 1,
\end{eqnarray}
we determine $M_{ij}^2$, 
the amplitude of $j$-plet component in $i$-th state as follows:
\begin{eqnarray}
M_{01}^2
&=&
\frac{1-g_1^2R^2}{1-(g_0g_1)^2}
\\
M_{08}^2
&=&
\frac{g_1^2(R^2-g_0^2)}{1-(g_0g_1)^2}
\\
M_{11}^2
&=&
\frac{g_0^2(1-g_1^2R^2)}{R^2\{1-(g_0g_1)^2\}}
\\
M_{18}^2
&=&
\frac{R^2-g_0^2}{R^2\{1-(g_0g_1)^2\}}
\end{eqnarray}
The amplitudes $M_{ji}^2$ 
obtained on the finest lattice
are plotted in Fig.~\ref{FcompoN}.
In the present quark-mass range,
ground (1st-excited) state is dominated by flavor-singlet (octet) components.

So far, we have argued the internal structures
from the view point of flavor contents.
It would be interesting and useful 
to interpret these flavor-mixing amplitudes in terms of a hadronic basis,
since hadronic-basis expansion of unstable particles
could be directly compared with experiments
and sometimes beneficial for model calculations.
Each flavor element can be decomposed into hadronic bases as
\begin{eqnarray}
|\Lambda_1\rangle
&=&
W_{N\bar K,1}|N\bar K\rangle
+
W_{\Sigma\pi,1}|\Sigma\pi\rangle
+ ....
\\
|\Lambda_8\rangle
&=&
W_{N\bar K,8}|N\bar K\rangle
+
W_{\Sigma\pi,8}|\Sigma\pi\rangle
+ ....
\end{eqnarray}
In this paper we restrict ourselves only to the $N\bar K$ and $\Sigma\pi$ bases.
The SU(3) Clebsh-Gordan coefficients tell us~\cite{Amsler:2008zzb}
\begin{eqnarray}
&&
| \Lambda_1 \rangle
\propto
\nonumber \\
&&
\frac{1}{\sqrt{8}}
\left(
\sqrt{2}|N\bar K\rangle
+
\sqrt{3}|\Sigma\pi\rangle
-
        |\Lambda\eta\rangle
-
\sqrt{2}|\Xi K\rangle
\right) \\
&&
| \Lambda_8 \rangle
\propto
\nonumber \\
&&
\frac{g_D}{\sqrt{20}}
\left(
\sqrt{ 2}|N\bar K\rangle
-
\sqrt{12}|\Sigma\pi\rangle
-
\sqrt{ 4}|\Lambda\eta\rangle
-
\sqrt{ 2}|\Xi K\rangle
\right)
\nonumber \\
&+&
\frac{g_F}{\sqrt{12}}
\left(
\sqrt{6}|N\bar K\rangle
+
\sqrt{6}|\Xi K\rangle
\right),
\end{eqnarray}
and the following expressions follows;
\begin{eqnarray}
W_{N\bar K,1}
&=&
\sqrt{\frac14} \\
W_{\Sigma\pi,1}
&=&
\sqrt{\frac38} \\
W_{N\bar K,8}
&=&
\frac{1}{\sqrt{g_D^2+g_F^2}}
\left(
\sqrt{\frac{1}{10}}g_D+\sqrt{\frac12}g_F
\right) \\
W_{\Sigma\pi,8}
&=&
\frac{1}{\sqrt{g_D^2+g_F^2}}
\left(
-\sqrt{\frac{3}{5}}g_D+\sqrt{\frac12}g_F
\right).
\end{eqnarray}
Here, $g_D$ and $g_F$ are unknown coupling constants
corresponding to two possible SU(3) constructions of the meson-baryon octet states,
${\rm Tr}\left(\left\{\bar B, B \right\} M\right)$
and
${\rm Tr}\left(\left[\bar B, B \right] M\right)$,
respectively.
We finally reach
the probabilities of $N\bar K$ and $\Sigma\pi$ states in $i$-th state,
$P_{N\bar K}^i$ and $P_{\Sigma\pi}^i$ ,
\begin{eqnarray}
P_{N\bar K}^i   &=& \left[\sum_{r=1,8}M_{ir}W_{N\bar K,r}  \right]^2
\\
P_{\Sigma\pi}^i &=& \left[\sum_{r=1,8}M_{ir}W_{\Sigma\pi,r}\right]^2.
\end{eqnarray}
Defining $\alpha$ by
\begin{equation}
\left(
\cos\alpha,\sin\alpha
\right)
\equiv
\left(
\frac{g_D}{\sqrt{g_D^2+g_F^2}},\frac{g_F}{\sqrt{g_D^2+g_F^2}}
\right)
\end{equation}
we obtain $P_{N\bar K}^i(\alpha)$ and $P_{\Sigma\pi}^i(\alpha)$ as functions of
the angle $\alpha$,
where we substitute 
$g_0$ and $g_1$ extrapolated to the chiral limit at $\beta =2.10$
(the finest lattice).
We simply adopt $R^2=1$, since $R^2$ does not deviate from unity,
and is almost quark-mass independent;
$R^2=$1.002(29), 1.015(21), 1.030(29), and 1.024(29) at $\kappa_{\rm
u,d}=$0.1357, 0.1367, 0.1374, and 0.1382 at this $\beta$.
The relative signs in $M_{ij}$ are determined
so that $M_{i1}M_{i8}(C)_{i1}(C)_{i2}>0$.

\begin{figure}[h]
\includegraphics[scale=1]{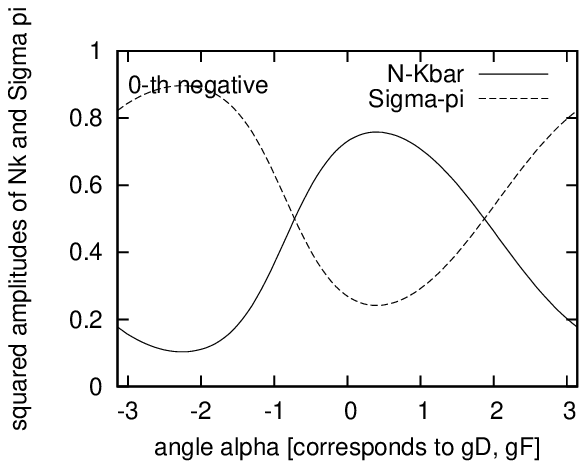}
\includegraphics[scale=1]{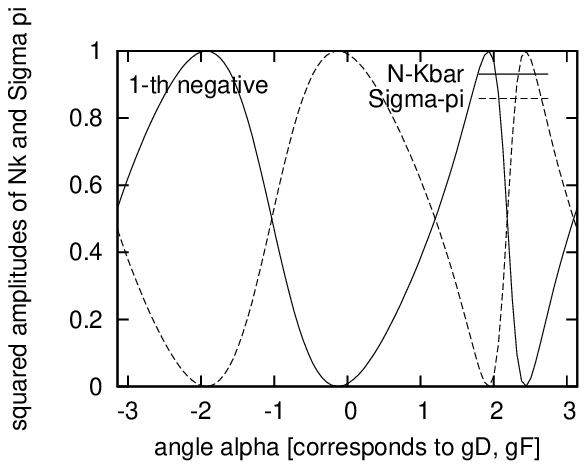}
\caption{\label{MBcompoN}
$P_{N\bar K}^i(\alpha)$ and $P_{\Sigma\pi}^i(\alpha)$ of the ground state (upper panel) and
of the 1st excited state (lower panel) as functions of $\alpha$.
They are determined from the $g_i^-$'s obtained at $\beta=2.10$
(the finest lattice).
}
\end{figure}

We focus on the meson-baryon components of the low-lying states
in negative-parity channel, since they are of interest here.
We show in Fig.~\ref{MBcompoN},
$P_{N\bar K}^i(\alpha)$ and $P_{\Sigma\pi}^i(\alpha)$ 
normalized as 
$P_{N\bar K}^i(\alpha) + P_{\Sigma\pi}^i(\alpha) = 1$ 
as a function of $\alpha$.
Though $g_i$'s are determined from those obtained at $\beta=2.10$,
the behaviors do not change significantly
even if we adopt extrapolated values on coarser lattices.
Now that the only unknown parameter is $\alpha$ representing $g_D$ and $g_F$
in the $J^P=1/2^-$ channel.
Unfortunately, 
the couplings, $g_D$ and $g_F$, should be determined up to 
including their signs
in order to estimate the meson-baryon components in each state
in this strategy. 
They can in principle be determined from lattice QCD computations
and their determination is left for forthcoming studies.

\section{Conclusion and Discussions}

An unquenched lattice QCD study for the low-lying $\Lambda$ baryon states
in the $S=-1$, $I=0$ and $J=1/2$ channel has been performed, focusing 
especially on the masses and structures of the two lowest negative-parity $\Lambda$'s.
We have constructed $2 \times 2$ cross correlators 
of the ``octet'' and ``singlet'' baryonic operators,
and diagonalized them so that we can properly extract the information of 
the low-lying two states.
In our measurements, we have adopted the 2-flavor gauge configurations
generated by CP-PACS collaboration with
the renormalization-group improved gauge action
and the ${\mathcal O}(a)$-improved quark action.
We have performed simulations at three different $\beta$'s,
$\beta = 1.80$, 1.95 and 2.10,
whose corresponding lattice spacings are $a = 0.2150$, 0.1555 and 0.1076
 fm, employing four different hopping parameters 
($\kappa_{\rm val}, \kappa_{\rm sea}$) for each cutoff, corresponding to 
the pion masses ranging from 500 MeV to 1.1 GeV.

It is shown that the mass of the ground-state $\Lambda(1115)$ is reproduced 
well at all three $\beta$'s, while the 1st-excited positive-parity $\Lambda$ state 
lies much higher than the experimentally observed $\Lambda(1600)$.
The same tendency was reported in Ref.~\cite{Burch:2006cc}, and 
also in the case of  the Roper resonance, $N^*(1440)$, 
which is the non-strange SU(3) partner of $\Lambda (1600)$~\cite{Sasaki:2001nf}.
It should also be noted that no $\pi\Sigma$ or $\bar K N$ scattering states 
are seen in the calculation.

Our results for negative-parity $\Lambda$ states indicate that 
there are two $1/2^-$ states nearly degenerate 
at around $1.6 -1.7$GeV,
while no state as low as $\Lambda (1405)$ is observed.
We have revealed the flavor structures of these states
from the lattice data for the first time.
It is found that 
the lowest negative-parity $\Lambda$ state is dominated by flavor-singlet component.
The second state, which is less than 100 MeV above the ground state, 
is predominantly flavor octet.  
Thus we find that the first two negative-parity $\Lambda$'s have different flavor
structures.

They, however, come significantly heavier than the experimentally observed
lowest-mass  state, $\Lambda(1405)$.
One naive possibility is that the obtained states do not correspond 
to the physical $\Lambda(1405)$, but
describe excited $\Lambda$ states.  In fact, the 2nd and 3rd negative-parity states lie at 1670 MeV
and 1800 MeV, both of which are three-star states in the Particle Data Group classification.
The present lattice data are consistent with these excited states. 
In the non-relativistic quark model approach, each of these states is classified as a flavor octet P-wave baryon.
We, however, have shown that the lower state is 
dominated by a flavor-singlet component.
So, the results here predict one flavor-singlet state 
and one flavor-octet state in the vicinity.
The reason for missing $\Lambda(1405)$ state will be 
its poor overlap with three-quark operators.
In fact, the inclusion of dynamical quarks does not
strongly enhance the signals of possible meson-baryon scattering states.
(See the later discussions.)
If $\Lambda (1405)$ is predominantly a meson-baryon molecular state,
such overlaps would be naturally small.

The other possibility is, of course, that
the lowest and the 2nd-lowest states describe physical 
$\Lambda(1405)$ and $\Lambda(1670)$
but the masses have been overestimated.
Considering that our simulation contains two-flavor dynamical quarks,
the failure of obtaining a light $\Lambda$ state could be attributed
either to
(1) strange-quark quenching,
(2) insufficient lattice volume or
(3) lack of chiral symmetry.
In fact, (1) seems to make the masses of octet baryons in positive
parity channel slightly ($\alt 10\%$) overestimated
in the present setups~\cite{AliKhan:2001tx}.
On the other hand, the deficiencies,(2) and (3), may cause the lowest
state not properly reproduced, supposing that the main component of
$\Lambda (1405)$ is a meson-baryon molecular state.
In order to check whether this conjecture is correct, simulations with
light dynamical quarks ($m_\pi \ll 500$ MeV)
and larger volume ($L \gg 2.5$ fm) will be required.

Upon the above conjecture,
we can further consider one interesting scenario that
these states both correspond to the physical $\Lambda(1405)$.
Then, the results may support its double pole structure proposed by the chiral unitary approach~\cite{Jido:2003cb}.
In our results, the lowest two states are almost degenerate at all the $\beta$'s (lattice spacing) and $\kappa$'s (quark masses).
Namely, the obtained two states are the signature of the double-pole resonance, but the
mass has not yet been reproduced because of the deficiency stated above.

\begin{figure}[hbt]
\includegraphics[scale=0.35]{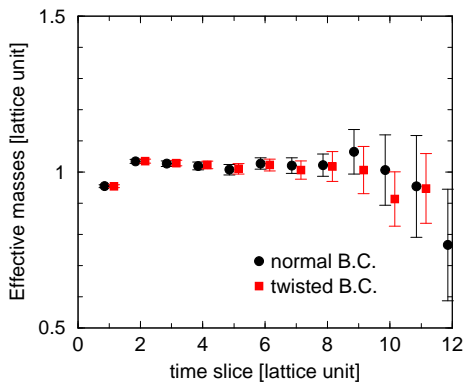}
\caption{\label{bccompare2}
Effective masses of the negative-parity $\Lambda$ ground state
under the normal (periodic) and the twisted boundary conditions.
}
\end{figure}

It is also important to note the missing meson-baryon scattering states.
Despite that dynamical up and down quarks are included
and the meson-baryon thresholds appear around/below the obtained eigen-energies,
we have found no clear signal of the meson-baryon scattering states.
This fact is supported by the observation that the effective mass plots
under the normal and twisted boundary conditions
show no prominent differences.
We show in Fig.\ref{bccompare2} the effective mass plots
of the ground state negative-parity $\Lambda$ under the normal and twisted b.c.'s.
They are obtained on the $24^3\times 48$ lattice with the largest
hopping parameter.
Thus we conclude that
no scattering states appear in the present spectrum.
Also in Ref.~\cite{Bulava:2009jb},
in which excited-state nucleon spectrum was systematically and extensively 
investigated with two-flavors of dynamical quarks,
no clear signal of scattering states was found
and the importance of multi-quark operators was raised.

A similar situation can be found in the computation of the Wilson loops,
whose expectation values give us the potential
between a (heavy fundamental) quark and an antiquark.
In the presence of dynamical quarks,
such an interquark potential should saturate and flatten at some interquark distance,
where the confining string is broken and a quark-antiquark pair is created.
However, no one has ever observed successfully such a ``string breaking'' effect
in the Wilson loop computations~\cite{Heller:1994rz,Bolder:2000un}.
One possible reason for this phenomenon is that
the Wilson loop itself has poor overlaps with such broken strings.

In the $\Lambda$ spectrum, we may similarly conjecture 
that the scattering states or broad resonances,
which are sensitive to boundary conditions,
have little overlap with the 3-quark interpolating field operators.
More detailed investigation would be needed
for clarification of 
scattering states purely induced by dynamical quarks.

The flavor structures of the $\Lambda$ states
have been very well clarified using the variational method.
The octet and the singlet components
are mixed when the flavor-SU(3) symmetry is broken.
Actually, the ground (1st-excited) state 
is dominated by singlet (octet) component, and 
the contamination by another representation
is at most 20\% (5\% when squared) in our present analysis.
From these findings, we expect that the flavor-SU(3) symmetry
is not largely broken.
A similar conclusion was also derived for the study of the meson-baryon coupling 
constants in lattice QCD~\cite{Erkol:2008yj}.

Because the SU(3) breaking effect seems small, the analyses without SU(3) mixings adopted so far
~\cite{Melnitchouk:2002eg,Nemoto:2003ft,Burch:2006cc,Ishii:2007ym}
make sense to some extent.
The mixings, however, get larger towards the chiral limit,
variational analyses could be essentially needed
when we adopt much lighter quarks.
We also find the meson-baryon contents in each state
strongly depend on the meson-baryon couplings and their signs.
Precise determination of the meson-baryon contents will require reliable determination of the couplings up to signs,
for which further lattice QCD calculations may be helpful.

\acknowledgments
All the numerical calculations were performed on NEC SX-8R at CMC, Osaka university and BlueGene/L at KEK. The unquenched gauge configurations employed in our analysis were all generated by CP-PACS collaboration~\cite{AliKhan:2001tx}. This work was supported in part by the Yukawa International Program for Quark-Hadron Sciences (YIPQS), by the Japanese Society for the Promotion of Science under contract number P-06327 and by KAKENHI (17070002, 19540275, 20028006 and 21740181).

\end{document}